\documentclass[fleqn,usenatbib,useAMS]{mnras}
\usepackage[T1]{fontenc}
\usepackage{ae,aecompl}
\usepackage{graphicx}
\usepackage{amsmath}
\usepackage{amssymb}
\newcommand{\bvec}[1]{\mbox{\boldmath $#1$}}

\title[PIC Simulations of the Filamentation Instability]{Kinetic Simulations of the
Filamentation Instability in Pair Plasmas}

\author[M. Iwamoto et al.]{
Masanori Iwamoto,$^{1,2}$\thanks{E-mail: iwamoto@esst.kyushu-u.ac.jp}
Emanuele Sobacchi,$^{3}$
and Lorenzo Sironi$^{2}$
\\
$^{1}$Faculty of Engineering Sciences, Kyushu University,
6-1, Kasuga-koen, Kasuga, Fukuoka, 816-8580, Japan\\
$^{2}$Department of Astronomy and Columbia Astrophysics Laboratory,
Columbia University, New York, NY 10027, USA\\
$^{3}$Racah Institute for Physics, The Hebrew University,
Jerusalem 91904, Israel
}

\date{Accepted XXX. Received YYY; in original form ZZZ}

\pubyear{2023}

\begin{document}
\label{firstpage}
\pagerange{\pageref{firstpage}--\pageref{lastpage}}

\maketitle

\begin{abstract}
  The nonlinear interaction between electromagnetic waves and plasmas
  attracts significant attention in astrophysics because it can affect
  the propagation  of Fast Radio Bursts (FRBs)---luminous
  millisecond-duration pulses detected at radio frequency.
  The filamentation instability (FI)--- a type of nonlinear
  wave-plasma interaction---is considered to be dominant near 
  FRB sources, and its nonlinear development may also affect 
  the inferred  dispersion measure of FRBs. 
  In this paper, we carry out fully kinetic particle-in-cell
  simulations of the FI in unmagnetized pair plasmas.
  Our simulations show that the FI generates transverse density filaments, 
  and that the electromagnetic wave propagates in near vacuum between them, as in a waveguide. 
  The density filaments keep merging until  force balance between the
  wave ponderomotive force and the plasma pressure gradient is established.
  We estimate the merging timescale and discuss the implications of
  filament merging for  FRB observations.
\end{abstract}

\begin{keywords}
plasmas -- instabilities -- relativistic processes -- fast radio bursts
\end{keywords}



\section{Introduction}\label{sec:intro}

The nonlinear interaction between electromagnetic waves and plasmas has
been widely studied in laboratory plasmas.
It is well-known that
the nonlinear interaction induces numerous plasma instabilities,
such as stimulated/induced Brillouin scattering (SBS),
stimulated/induced Raman scattering, filamentation instability (FI),
modulation instability, two-plasmon decay instability, and oscillating
two-stream instability
\citep[e.g.,][]{Kaw1973,Max1973a,Max1974,Drake1974,Forslund1975,Mima1975,Mima1984,
Cohen1979, Kruer1988}.
The SBS is also referred to as induced Compton scattering when 
kinetic effects are important.
These nonlinear phenomena play a crucial role for various laser-plasma
experiments, like wakefield acceleration \citep{Tajima1979} and fast ignition
of inertial confinement fusion \citep{Tabak1994,Deutsch1996}.

Recently, the nonlinear wave-plasma interaction has attracted significant attention
from astrophysics in the context of Fast Radio Bursts (FRBs).
FRBs are extremely bright millisecond duration pulses at radio frequency
and often show a high degree of linear polarization
\citep[e.g.,][]{Lorimer2007,Michilli2018,Day2020,Luo2020,Nimmo2021}.
Magnetars have emerged as one of the leading FRB progenitors
\citep[e.g.,][]{Andersen2020,Bochenek2020,Lyubarsky2021}.
In the magnetar scenario, the FRB radio pulse propagates through the magnetar wind, which consists of a
pair (electron-positron) plasma.
The stimulated/induced Raman scattering, two-plasmon decay instability,
oscillating two-stream instability, and modulation instability do not occur
for linearly polarized pump waves propagating through pair plasmas because
of the lack of electrostatic plasma waves \cite[cf.,][]{Matsukiyo2003}.
Therefore, only the SBS and FI can operate near FRB progenitors.
Recently, \cite{Ghosh2022} demonstrated that
the SBS is suppressed for realistic pump waves with a broad
spectrum and the FI is then the prevailing process.
On the other hand, the development of the FI can profoundly affect the wave propagation.
\cite{Sobacchi2022b} pointed out that the FI generates transverse density filaments separated by near-vacuum regions.
The FRB waves propagate in the near-vacuum regions like in a waveguide, and this can significantly affect the inferred dispersion measure of FRBs.
The FI must be taken into account for the propagation  of the FRB
radio pulses.

The excitation of the FI is confirmed by  particle-in-cell (PIC)
simulations of relativistic magnetized shocks
\citep{Iwamoto2017, Iwamoto2022, Plotnikov2018, Babul2020, Sironi2021},
in which the electromagnetic waves are excited self-consistently in the shock transition.
Relativistic magnetized shocks are often
considered to be one of the candidates for the origin of the coherent
FRB emission
\cite[e.g.,][]{Lyubarsky2014,Beloborodov2017,Beloborodov2020,Metzger2019,
Plotnikov2019,Margalit2020a,Margalit2020b}.
The wave emission from the shock front is very strong, in
the sense that the wave strength parameter is much greater than unity,
$a_0= eE_0/m_ec\omega_0 \gg 1$ \citep{Iwamoto2017},
where $E_0$ is the wave amplitude and $\omega_0$ is the wave frequency,
indicating that the radio pulses satisfy $a_0 \gg 1$
in the vicinity of the FRB progenitors
\citep[see, e.g.,][]{Beloborodov2020}.
Although the wave amplitude drastically decreases with distance from the
sources, the previous studies \citep{Sobacchi2022a, Sobacchi2022b} showed that the FI
has significant influence on the propagation process of the radio pulses even
for $a_0 \ll 1$. In this paper, we focus on the regime $a_0 \ll 1$ in which
the radio pulses are far away from the sources.

The FI is caused by the ponderomotive force, which
expels particles from the regions of high wave intensity.
The refractive index increases in the low density region, where the
electromagnetic waves are in turn accumulated and the wave intensity is
further enhanced, completing the feedback loop.
The plasma temperature in the resulting high density region gradually
increases due to  adiabatic heating and
this loop ceases---equivalently, the instability saturates---when  force balance between
the wave ponderomotive force and the plasma pressure gradient is achieved
\citep{Kaw1973,Sobacchi2022b}.
When the initial particle thermal energy $m_ec^2\beta_{th0}^2$ is much
smaller than the pump wave ponderomotive potential $m_ec^2a_0^2/4$,
\begin{equation}
    \label{eq:cold}
    \beta_{th0} \ll a_0,
\end{equation}
a high density compression is required for the force
balance and so the density fluctuation achieves substantial amplitudes. 
Here, $\beta_{th0}$ is the thermal velocity normalized by the speed of light
$c$. 
Therefore, the FI leads to a significant density contrast for
$\beta_{th0} \ll a_0$, a condition which can be satisfied in FRB environments
\citep{Sobacchi2022b}.

The plasma temperature plays an important role for the linear
evolution of the FI as well.
It is well-known that the linear growth rate  transitions
from weak to strong coupling
\citep[e.g.,][]{Drake1974,Forslund1975,Cohen1979,Kruer1988}.
In the strong coupling regime, the nonlinear effect is quite significant
and the density fluctuation is no longer a normal mode of the plasma.
Considering the cold plasma condition (Equation \ref{eq:cold}),
we obtain the threshold for the weak and strong coupling regimes
(see Section \ref{sec:linear} for the detailed derivation),
respectively,
\begin{gather}
\label{eq:weak}
\sqrt{a_0\frac{\omega_{pe}}{\omega_0}}  \ll \beta_s \ll a_0 \ {\rm (weak \ coupling)}, \\
\label{eq:strong}
\beta_s \ll \sqrt{a_0\frac{\omega_{pe}}{\omega_0}} \ {\rm (strong \ coupling)},
\end{gather}
where $\beta_s$ is the sound speed normalized by the speed of light
and $\omega_{pe}$ is the plasma frequency. Here we have assumed
the limit of a high frequency pump wave with $\omega_{0} \gg \omega_{pe}/a_0$,
which is valid for FRB environments \citep{Sobacchi2022b}.
In the strong (respectively, weak) coupling regime the e-folding time of the FI is shorter (respectively, longer) than the sound crossing time of the density filaments, 
as discussed in Section \ref{sec:linear}.
We investigate the FI for these two cases.

In this paper, we perform PIC simulations and study the FI
in pair plasmas, a composition which is still under-explored
because laboratory plasmas are generally ion-electron plasmas.
Although \cite{Ghosh2022} carried out  PIC simulations of the FI in pair plasmas, 
they focused on the linear phase. We follow the long-term evolution of the FI and 
discuss the saturation mechanism in more detail.
This paper is organized as follows.
We reproduce the linear analysis of the FI for the sake of completeness
in Section \ref{sec:linear}.
Section \ref{sec:simulation} describes our simulation results.
We compare them with the linear analysis and describe the
saturation mechanism of the FI.
In Section \ref{sec:sum}, we summarize this study and
discuss its implications for FRBs.

\section{Linear Analysis}\label{sec:linear}

We here reproduce the linear growth rate of the FI
for the sake of completeness. This linear analysis is based on previous
works \citep{Edwards2016,Schluck2017,Sobacchi2020,Sobacchi2022a,Ghosh2022}.

\subsection{Fluid Approximation}\label{ssec:flu}

The linearly polarized electromagnetic pump wave is described by the wave 
equation,
\begin{gather}
  \left[\Delta-\frac{1}{c^2}\frac{\partial^2}{\partial t^2}\right]\bvec{A}
  =-\frac{4\pi}{c}\bvec{J},
\end{gather}
where the Coulomb gauge condition $\nabla \cdot \bvec{A} = 0$ is applied.
Let us assume an unmagnetized pair plasma governed by
fluid equations,
\begin{gather}
  \frac{\partial}{\partial t} (\gamma_jn_j) +
  \bvec{\nabla} \cdot (\gamma_jn_j\bvec{v_j})= 0, \\
  \frac{\partial}{\partial t}(\gamma_j\bvec{v_j}) + (\bvec{v_j} \cdot
  \bvec{\nabla})\gamma_j\bvec{v_j} = \nonumber \\
  -c_s^2\frac{\bvec{\nabla}n_j}{\gamma_jn_j}+\frac{q_j}{m_jc}
  \left[-\frac{\partial \bvec{A}}{\partial t}
  + \bvec{v_j} \times (\bvec{\nabla} \times \bvec{A})\right],\\
  \bvec{J} = \sum_{j}q_jn_j\bvec{v_j},
\end{gather}
where the subscript $j=e,p$ represents particle species
(i.e., electron and positron) and $\gamma_j$ is the Lorentz factor.
We assume that the electron temperature is equal to the positron one
and non-relativistic $c_s \ll c$.
The vector potential of the pump wave $\bvec{A_0}$ is
given by
\begin{equation}
  \bvec{A_0} =(0,A_0\sin\phi_0,0)
\end{equation}
where $\phi_0=k_0x-\omega_0t$.
We assume that the wave frequency $\omega_0$
is much higher than the electron plasma frequency
$\omega_{pe}=\sqrt{4\pi n_0e^2/m_e}$ (i.e., $\omega_0 \simeq ck_0$),
where $n_0$ is the unperturbed electron density and $n_e=n_p=n_0$ is 
initially satisfied. 
The wave amplitude is small in the sense that
the wave strength parameter $a_0$ is sufficiently smaller than unity,
\begin{equation}
  a_0 = \frac{eA_0}{m_ec} \ll 1.
\end{equation}
By substituting $\bvec{A_0}$ into the basic equations,
we obtain the zeroth-order three velocity $\bvec{v_0}$ and density $n_0+\delta n_0$,
\begin{align}
  v_{0jx} &= \frac{1}{4}ca_0^2(1-\cos2\phi_0), \\
  v_{0jy} &= \pm ca_0\sin\phi_0\left(1-\frac{1}{4}a_0^2+\frac{1}{4}a_0^2\cos2\phi_0\right),\\
  \label{eq:dn0}
  \delta n_0 &= -\frac{1}{4}n_0a_0^2\cos2\phi_0,
\end{align}
where the positive (negative) sign corresponds to the electron (positron).
The dispersion relation including the lowest-order nonlinear correction
is \citep[e.g.,][]{Sluijter1965, Max1974}
\begin{equation}
  \omega_0^2 - c^2k_0^2-2\omega_{pe}^2\left(1-\frac{1}{4}a_0^2\right)=0.
\end{equation}
Although the zeroth-order solution is valid only for weak,
high-frequency electromagnetic waves
and does not represent an exact steady-state solution, which can not
be analytically derived \citep[see, e.g.,][]{Kaw1970,Max1973b},
we now perturb this quasi-equilibrium and study the nonlinear interaction
between the pump wave and the unmagnetized pair plasma.
Considering only the lowest-order coupling
$(\omega_{\pm}, \bvec{k_{\pm}}) = (\omega_0 \pm \omega, \bvec{k_0} \pm
\bvec{k})$, which is valid for $a_0 \ll 1$, the perturbed quantities are
written as
\begin{align}
  \bvec{A} &=  \bvec{A_0} +\bvec{\delta A_+}e^{i\phi_+} + \bvec{\delta A_-}e^{i\phi_-}+c.c.\\
  \bvec{v_e} &= \bvec{v_{0e}}+\bvec{\delta v}e^{i\phi} + \bvec{\delta v_{+}}e^{i\phi_+} +
  \bvec{\delta v_{-}}e^{i\phi_-}+c.c., \\
  \bvec{v_p} &= \bvec{v_{0p}}+\bvec{\delta v}e^{i\phi} - \bvec{\delta v_{+}}e^{i\phi_+} -
  \bvec{\delta v_{-}}e^{i\phi_-}+c.c., \\
  n_e &= n_0+\delta n_0+\delta n e^{i\phi} + c.c., \\
  n_p &= n_0+\delta n_0+\delta n e^{i\phi} + c.c..
\end{align}
where $c.c.$ indicates the complex conjugate,
$\phi = \bvec{k}\cdot \bvec{x} -\omega t$, $|\omega| \ll \omega_0$, and
$\phi_{\pm} = \bvec{k_{\pm}}\cdot \bvec{x} -\omega_{\pm} t = \phi_0 \pm \phi$.
We assume that no charge separation is excited,
which is valid for a linear polarized pump wave \citep[cf.,][]{Matsukiyo2003}.
Substituting these into the linearized equations
and neglecting the non-resonant terms
$\propto e^{i( 2\phi_0 \pm \phi)}, e^{i( 3\phi_0 \pm \phi)}$,
we finally obtain the dispersion relation,
\begin{equation}
  \label{eq:gamf}
  \frac{1}{2}a_0^2\omega_{pe}^2(Q_{fluid}-1)\left( \frac{\cos^2\theta_+}{D_+}+\frac{\cos^2\theta_-}{D_-}\right)=1,
\end{equation}
where
\begin{gather}
  \cos\theta_\pm = \frac{\bvec{A_0}\cdot \bvec{\delta A_\pm}}
  {|\bvec{A_0}| |\bvec{\delta A_\pm}|},\\
  Q_{fluid} = \frac{c^2k^2}{D_a}, \\
  D_\pm = \omega_\pm^2 - c^2k_\pm^2-2\omega_{pe}^2\left(1-\frac{1}{4}a_0^2\right), \\
  D_a = \omega^2-c_s^2k^2.
\end{gather}
$D_\pm=0$ and $D_a=0$ describe the dispersion relation of the scattered
electromagnetic waves and sound waves, respectively.
We here assume that the scattering occurs only in the $x-y$ plane
(i.e., $\bvec{\delta A_\pm}$ lies in the $x-y$ plane).
Considering $\bvec{\bvec{k_0}} \perp \bvec{A_0}$ and
$\bvec{\bvec{k_\pm}} \perp \bvec{\delta A_\pm}$,
$\cos\theta_\pm $ satisfies
\begin{equation}
  \cos\theta_\pm = \frac{\bvec{k_0}\cdot \bvec{k_\pm}}
  {|\bvec{k_0}| |\bvec{k_\pm}|}.
\end{equation}

\begin{figure}
  \begin{center}
    \includegraphics[width=7cm]{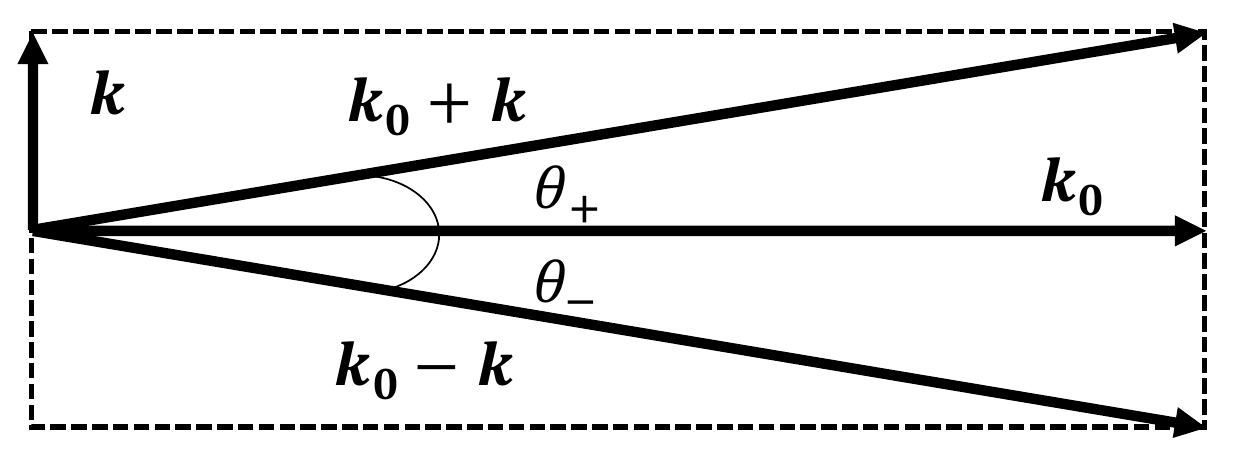}
  \end{center}
\caption{Wavevector diagram for the FI.}
\label{fig:vec}
\end{figure}

The FI can be interpreted as the four-wave coupling
\citep[e.g.,][]{Drake1974,Kruer1988},
\begin{equation}
  \label{eq:fif}
  D_+=D_-=0.
\end{equation}
Equation \ref{eq:fif} can be satisfied only for $k \ll k_0$,
showing that the FI originates from two forward-scattered electromagnetic
waves. The wavevector geometry of the FI is sketched in Figure \ref{fig:vec}.
We can evaluate the real frequency of the FI from Equation \ref{eq:fif},
\begin{equation}
  {\rm Re} (\omega) = \frac{c^2\bvec{k_0}\cdot\bvec{k}}{\omega_0},
\end{equation}
where $c^2\bvec{k_0}/\omega_0$ is the group velocity of the pump wave.
Since ${\rm Re} (\omega) \sim 0$ is satisfied for
$\bvec{k_0} \cdot \bvec{k} \sim 0$, the FI is a purely growing mode.

We now estimate the maximum growth rate of the FI.
For the FI, we can safely assume $\bvec{k} \cdot \bvec{k_0} \sim 0$ and
$\cos\theta_{\pm} \sim 1$. For $|\omega| \ll ck$,
Equation \ref{eq:gamf} reduces to
\begin{equation}
  \label{eq:gamf3}
  \left(\omega^2-c_s^2k^2\right)
  \left(\omega^2-\frac{c^4k^4}{4\omega_0^2}\right)
  =\frac{a_0^2\omega_{pe}^2c^4k^4}{4\omega_0^2}.
\end{equation}
Substituting $\omega = i \Gamma$, where $\Gamma \ll c_sk$, 
into Equation \ref{eq:gamf3}, we obtain
\begin{equation}
  \Gamma^2+\frac{c^4k^4}{4\omega_0^2}=\frac{a_0^2\omega_{pe}^2c^2k^2}{4\omega_0^2\beta_s^2}.
\end{equation}
The condition $\Gamma \ll c_sk$ is generally referred to as the weak coupling 
regime \citep[e.g.,][]{Drake1974,Forslund1975,Cohen1979,Kruer1988}.
We can find the maximum growth rate and corresponding wavevector,
\begin{gather}
  \label{eq:gamw}
  \Gamma_{max} = \frac{a_0^2\omega_{pe}^2}{4\beta_s^2\omega_0} \ {\rm (weak \ coupling)}, \\
  \label{eq:kyw}
  k_y = \frac{a_0\omega_{pe}}{\sqrt{2}c\beta_s} \ {\rm (weak \ coupling)}.
\end{gather}
The validity condition $\Gamma \ll c_sk$ is
\begin{equation}
  \label{eq:vtew}
  \beta_s \gg \sqrt{a_0\frac{\omega_{pe}}{\omega_0}} \ {\rm (weak \ coupling)}.
\end{equation}
For $\Gamma \gg c_sk$, which is the so-called strong coupling regime,
Equation \ref{eq:gamf3} reduces to
\begin{equation}
  \Gamma^4+\frac{c^4k^4}{4\omega_0^2}\Gamma^2
  =\frac{a_0^2\omega_{pe}^2c^4k^4}{4\omega_0^2}.
\end{equation}
The growth rate increases with $k$ and
the asymptotic solution is written as
\begin{equation}
  \label{eq:gams}
  \Gamma_{max} = a_0\omega_{pe} \ {\rm (strong \ coupling)}.
\end{equation}
$\Gamma$ is then expanded for large $k$,
\begin{equation}
  \Gamma=
  \left(1-\frac{4a_0^2\omega_0^2\omega_{pe}^2}{c^4k^4}\right)\Gamma_{max}.
\end{equation}
Thus $\Gamma$ asymptotically approaches the maximum for
\begin{equation}
  k \gg \frac{\sqrt{a_0\omega_0\omega_{pe}}}{c}.
\end{equation}
We have neglected factors of order of unity.
The validity condition is
\begin{gather}
  \label{eq:kys}
  \frac{\sqrt{a_0\omega_0\omega_{pe}}}{c} \ll k_y 
  \ll \frac{a_0\omega_{pe}}{c\beta_s} \ {\rm (strong \ coupling)} ,\\
  \label{eq:vtes}
  \beta_s \ll \sqrt{a_0\frac{\omega_{pe}}{\omega_0}} \ {\rm (strong \ coupling)}.
\end{gather}
This condition and maximum growth rate
show that the e-folding time of the FI
$\tau_{grow} \sim 1/\Gamma_{max} \sim 1/a_0\omega_{pe}$ is much
shorter than the sound crossing time of the density filaments
$\tau_{cross} \sim 1/c_sk_y \gg 1/a_0\omega_{pe}$ for the strong coupling
regime. On the other hand,
$\tau_{cross}/\tau_{grow} \sim a_0\omega_{pe}/\beta_s^2\omega_0 \ll 1$
is satisfied for the weak coupling regime. This difference affects
the heating physics during the linear and nonlinear evolution of the FI
(see Section \ref{sec:sat}).

\subsection{Fully Kinetic Formulation}

We here assume an unmagnetized pair plasma governed
by the Vlasov equation,
\begin{gather}
  \frac{\partial f_j}{\partial t}+ \bvec{v_j} \cdot
  \frac{\partial f_j}{\partial \bvec{x}}
  +\frac{q_j}{m_jc}\left[-\frac{\partial \bvec{A}}{\partial t}
  +\bvec{v_j} \times (\bvec{\nabla} \times \bvec{A})\right] \cdot
  \frac{\partial f_j}{\partial \bvec{u}}=0,\\
  \bvec{J} = \sum_{j}q_j \int \bvec{v}f_j{\rm d}\bvec{u},
\end{gather}
where $\bvec{u=\gamma \bvec{v}}$ is the particle four velocity.
Let us assume that the zeroth-order distribution function $f_{0j}$
satisfies
\begin{equation}
  \int f_{0j}{\rm d}\bvec{u} = n_0\left(1-\frac{1}{4}a_0^2\cos2\phi_0\right),
\end{equation}
which is motivated by the fluid approximation in Equation \ref{eq:dn0}.
$f_{0j}$ is then written as
\begin{equation}
  f_{0j} = n_0\left(1-\frac{1}{4}a_0^2\cos2\phi_0\right)F_0(\bvec{u_{j \perp}})\delta\left(\bvec{u_{j \parallel}}+\frac{q_j\bvec{A}}{m_j}\right),
\end{equation}
where $\bvec{u_{j \parallel}}$ and $\bvec{u_{j \perp}}$
are the four velocity components of parallel and perpendicular to the vector
potential $\bvec{A}$, respectively.
The $\delta$ is the Dirac delta function and this term comes from the
conservation of the canonical momentum.
For $a_0 \ll 1$, $F_0$ is given by the
non-relativistic 1D Maxwellian distribution,
\begin{equation}
  F_0 = \frac{1}{\sqrt{2\pi}v_{th0}}\exp\left[-\frac{v_x^2}{2v_{th0}^2}\right],
\end{equation}
where $v_{th0}=\sqrt{k_BT_e/m_e}$ is the thermal velocity
and the electron temperature $T_e$ is equal to the positron one $T_p$.
Substituting $\bvec{A_0}$ and $f_{0j}$ into the Vlasov and wave equations,
we obtain the dispersion relation
\begin{equation}
  \omega_0^2 - c^2k_0^2-2\omega_{pe}^2\left(1-\frac{1}{4}a_0^2\right)=0,
\end{equation}
which is identical to the fluid approximation.
Considering only the lowest-order coupling, which is valid for $a_0 \ll 1$,
the perturbed quantities can be expressed as
\begin{gather}
  \bvec{A} =  \bvec{A_0} + \bvec{\delta A_+}e^{i\phi_+} + \bvec{\delta A_-}e^{i\phi_-}+c.c.,\\
  f_j = n_0\left(1-\frac{1}{4}a_0^2\cos2\phi_0\right)
  (F_0+\delta Fe^{i\phi})
  \delta\left(\bvec{u_{j\parallel}}+\frac{q_j\bvec{A}}{m_j}\right)
  +c.c.,
\end{gather}
where $\delta F$ is independent of $\bvec{u_{j \parallel}}$.
Linearizing the basic equations, we finally obtain the fully kinetic 
dispersion relation,
\begin{equation}
  \label{eq:gamk}
  \frac{1}{2}a_0^2\omega_{pe}^2(Q_{kin}-1)\left( \frac{\cos^2\theta_+}{D_+}+\frac{\cos^2\theta_-}{D_-}\right)=1,
\end{equation}
where
\begin{gather}
  Q_{kin} = \frac{c^2}{2v_{th0}^2}\frac{{\rm d}Z}{{\rm d}\zeta}, \\
  \zeta = \frac{\omega}{\sqrt{2}v_{th0}k}.
\end{gather}
$Z(\zeta)$ is the plasma dispersion function given by
\begin{gather}
  Z(\zeta) = \frac{1}{\sqrt{\pi}}\int_{-\infty}^{\infty} \frac{e^{-z^2}}{z-\zeta}{\rm d}z, \\
  \frac{{\rm d}Z}{{\rm d}\zeta} = -2(1+\zeta Z).
\end{gather}
The difference from the fluid approximation is that the sound wave
dispersion relation in Equation \ref{eq:gamf} is replaced by the kinetic one.

\begin{figure*}
 \includegraphics[width=8cm]{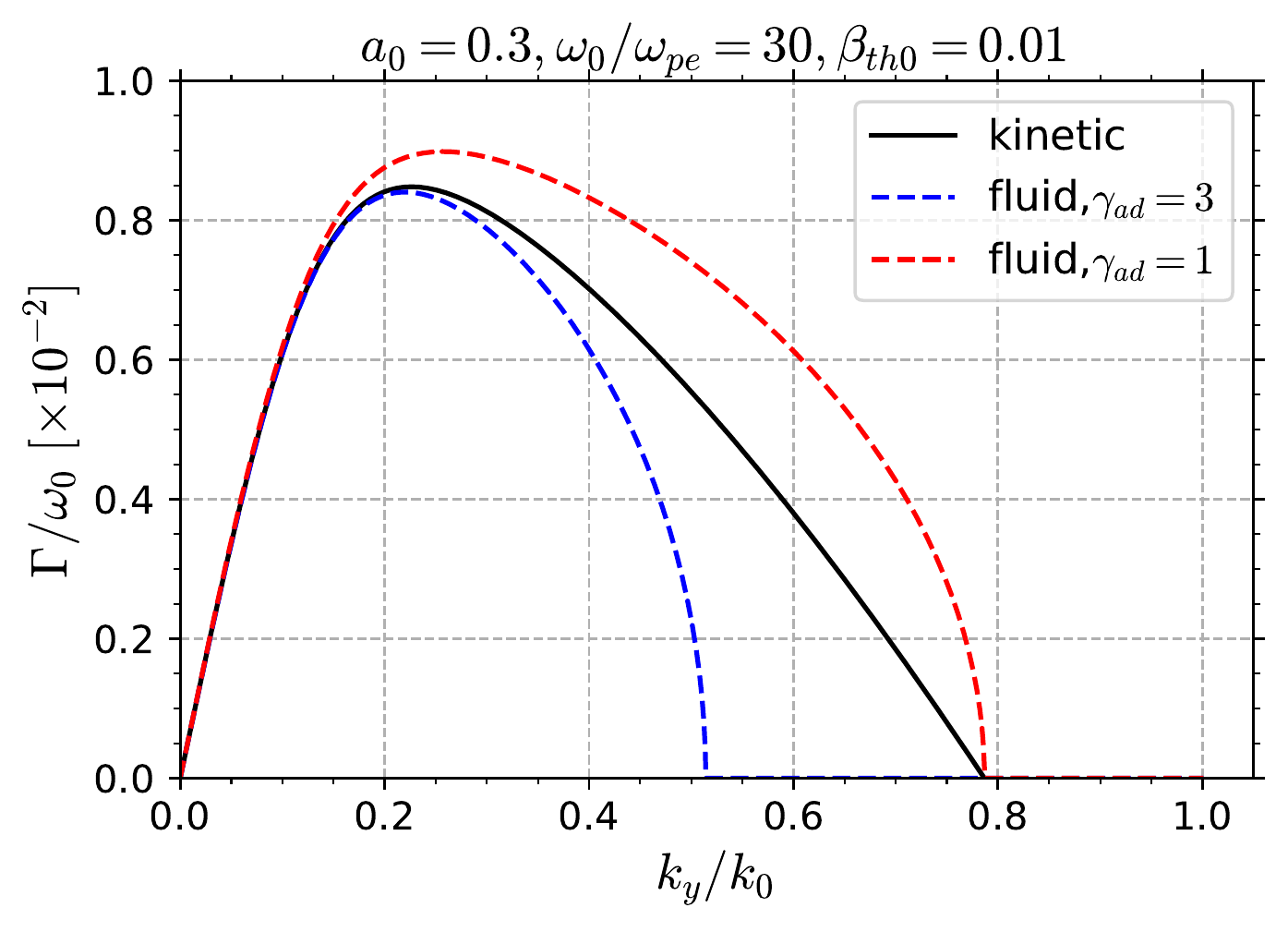}
 \includegraphics[width=8cm]{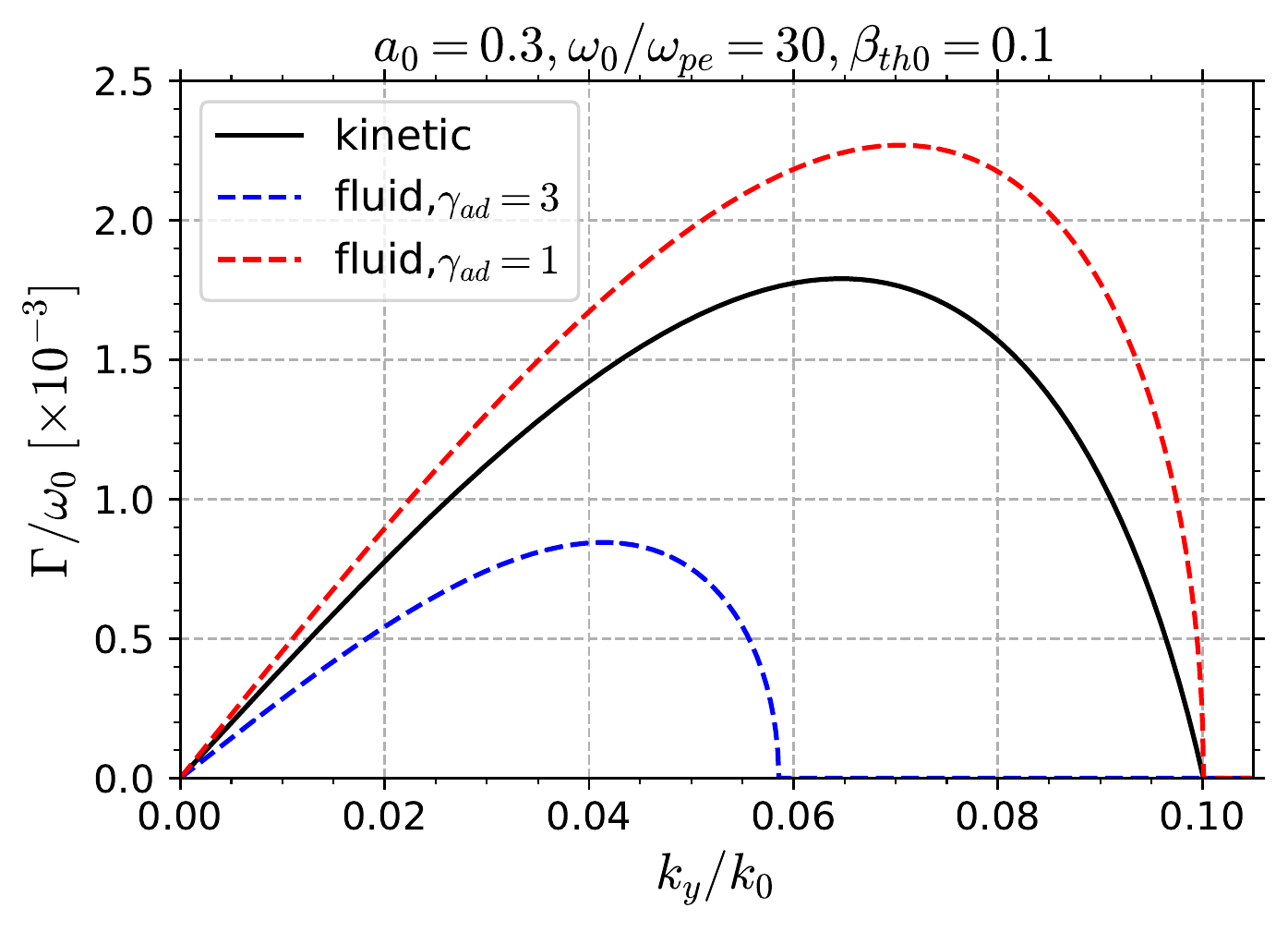}
\caption{Linear growth rate of the FI for the strong
(left) and weak (right) coupling cases.}
\label{fig:linear}
\end{figure*}

We numerically derive the linear growth rate of the FI and show it 
in Figure \ref{fig:linear} for
$(a_0,\omega_0/\omega_{pe},\beta_{th0})=(0.3,30,0.01)$ (left)
and $(0.3,30,0.1)$ (right).
Our simulations are performed for these two cases.
The black solid lines in Figure \ref{fig:linear} indicate the kinetic growth rates.
We also show the fluid ones with the adiabatic index
$\gamma_{ad}=1$ (isothermal) and $\gamma_{ad}=3$ (1D gas) in red and
blue dashed lines, respectively.
The left panel refers to the strong coupling regime $|\omega| \gg c_sk$.
Since the results of fluid and kinetic calculations
are comparable as further discussed below, we can safely
use the analytical estimates from the fluid approximation
and Equation \ref{eq:gams} and \ref{eq:kys} give, for the strong coupling regime,
\begin{gather}
  \frac{\Gamma_{max}}{\omega_0} \sim 1.0 \times 10^{-2}, \\
  0.1 \ll \frac{k_y}{k_0} \ll  1.
\end{gather}
In contrast, for the weak coupling regime $|\omega| \ll c_sk$ in the right panel,
the maximum growth rate and the corresponding wavevector are
estimated from Equation \ref{eq:gamw} and \ref{eq:kyw},
\begin{gather}
  \frac{\Gamma_{max}}{\omega_0} \sim 2.5 \times 10^{-3}, \\
  \frac{k_y}{k_0} \sim 0.07.
\end{gather}
Here we have assumed $\beta_s \sim \beta_{th0}$.
These analytical estimates are roughly consistent with the numerical results.

We now expand why the fluid and kinetic calculations
give comparable results.
This is not surprising because
the density fluctuation is a non-propagating mode
and the FI is almost unaffected by the Landau damping as already
discussed by \cite{Cohen1979}.
The derivative of the plasma dispersion function is
expressed by the expansion \citep[see, e.g.,][]{Fried1961}
for $|\zeta| \gg 1$ (i.e., strong coupling regime $|\omega| \gg c_sk$),
\begin{equation}
  \frac{{\rm d}Z}{{\rm d}\zeta} =
  \frac{1}{\zeta^2}+\frac{3}{2}\frac{1}{\zeta^4}+
  \frac{15}{4}\frac{1}{\zeta^6}+\cdots,
\end{equation}
and for $|\zeta| \ll 1$,
\begin{equation}
  \frac{{\rm d}Z}{{\rm d}\zeta}=
  -2\sqrt{\pi}i\zeta e^{-\zeta^2}
  -2+4\zeta^2-\frac{8}{3}\zeta^4+\cdots.
\end{equation}
Here we have used ${\rm Im}(\zeta)>0$.
$Q_{kin}$ is thus approximately expressed as for the strong coupling regime
$|\omega| \gg c_sk$,
\begin{equation}
  Q_{kin} \simeq
  \frac{c^2k^2}{\omega^2}\left(1+\frac{3v_{th0}^2k^2}{\omega^2}\right).
\end{equation}
and for the weak coupling regime $|\omega| \ll c_sk$,
\begin{equation}
  Q_{kin} \simeq -\frac{c^2}{v_{th0}^2}.
\end{equation}
$Q_{fluid}$ is expressed as for $|\omega| \gg c_sk$,
\begin{equation}
  Q_{fluid} \simeq\frac{c^2k^2}{\omega^2}\left(1+\frac{c_s^2k^2}{\omega^2}\right).
\end{equation}
and for $|\omega| \ll c_sk$,
\begin{equation}
  Q_{fluid} \simeq -\frac{c^2}{c_s^2}.
\end{equation}
If we assume the adiabatic index
$\gamma_{ad}=3$ for $|\omega| \gg c_sk$ and
$\gamma_{ad}=1$ for $|\omega| \ll c_sk$,
$Q_{kin}$ is identical to $Q_{fluid}$.
Therefore, the fluid approximation for the FI is reasonable.

On the other hand, the effect of the Landau damping is not
negligible for the SBS because the SBS induces sound-like waves which
are heavily damped unless a strong temperature difference between electrons and positrons is induced.
The fluid approximation for the SBS is then valid only for the strong coupling
regime (see Appendix \ref{app:sbs}).

\section{Numerical Simulation}\label{sec:simulation}

\subsection{Setup}

We use a fully kinetic particle-in-cell (PIC) code
\citep{Matsumoto2015,Matsumoto2017}, which employs
an implicit Maxwell solver without any digital filters \citep{Ikeya2015},
a charge conservation scheme for the electric current deposition
\citep{Esirkepov2001}, and a second-order shape function
for computational macroparticles.
We consider a rectangular simulation box in $x$-$y$ plane
and the boundary condition in all directions is
periodic for both the fields and the particles.
All three components of fields and velocities are tracked in
our simulations.
The initial condition is based on \cite{Ghosh2022}.
The plane monochromatic pump wave is initially introduced,
\begin{align}
  \bvec{E_0} &= (0,E_0\cos{k_0x},0), \\
  \bvec{B_0} &= \left(0,0,\frac{ck_0}{\omega_0}E_0\cos{k_0x} \right).
\end{align}
We also study the case of a pump wave vector potential perpendicular to
simulation plane (see Appendix \ref{app:out}).
This pump wave propagates through
homogeneous, unmagnetized pair plasmas with a Maxwellian distribution.
We calculate the initial thermal spread $\beta_{th0}$ in the proper frame.
The initial bulk four velocity satisfies
\begin{align}
  \bar{u}_{0jx} &= \frac{1}{2}ca_0^2\sin^2{k_0x}, \\
  \bar{u}_{0jy} &= \pm ca_0\sin{k_0x}, \\
  \bar{u}_{0jz} &= 0,
\end{align}
where the positive (negative) sign corresponds to the electron (positron).
The SBS generally grows faster than the FI for monochromatic pump waves
\citep{Ghosh2022}.
The simulation domain in the $x$ direction
is just one wavelength of the pump wave $L_x = \lambda_0$,
where $\lambda_0$ is the wavelength of the pump wave.
Since the backward SBS is most unstable
and the wavenumber of the back-scattered wave can
be estimated as
$\bvec{k_s} = \bvec{k_0}-\bvec{k} \simeq -(1-2\beta_{s})k_0\bvec{\hat{x}}$
\citep[e.g.,][]{Kruer1988}, the SBS can be suppressed by a small box
as already discussed by \cite{Ghosh2022}.
This is the case for the weak coupling case,
however, the SBS grows into an substantial amplitude for
the strong coupling case (see Appendix \ref{app:sbs}).
The simulation domain in the $y$ direction is
$L_y = 120\lambda_0 = 8 \pi c/\omega_{pe}$
to follow the filament mergers.
The grid size and time step are respectively set as
$\Delta x/\lambda_0= 0.005$ and  $\omega_0 \Delta t = 0.0314$.
The number of particles per cell per species is $n_0 \Delta x^2 = 32$.
Tests of numerical convergence are shown in Appendix \ref{app:conv}.

We fix the pump wave frequency $\omega_0/\omega_{pe} = 30$ and
the wave strength parameter $a_0=0.3$ throughout this study.
We carry out our simulations for strong and weak coupling cases: 
$\beta_{th0}=0.01$ and $0.1$, which satisfy the condition \ref{eq:cold}. 

\subsection{Simulation Results}

\begin{figure*}
\begin{center}
  \includegraphics[width=8cm]{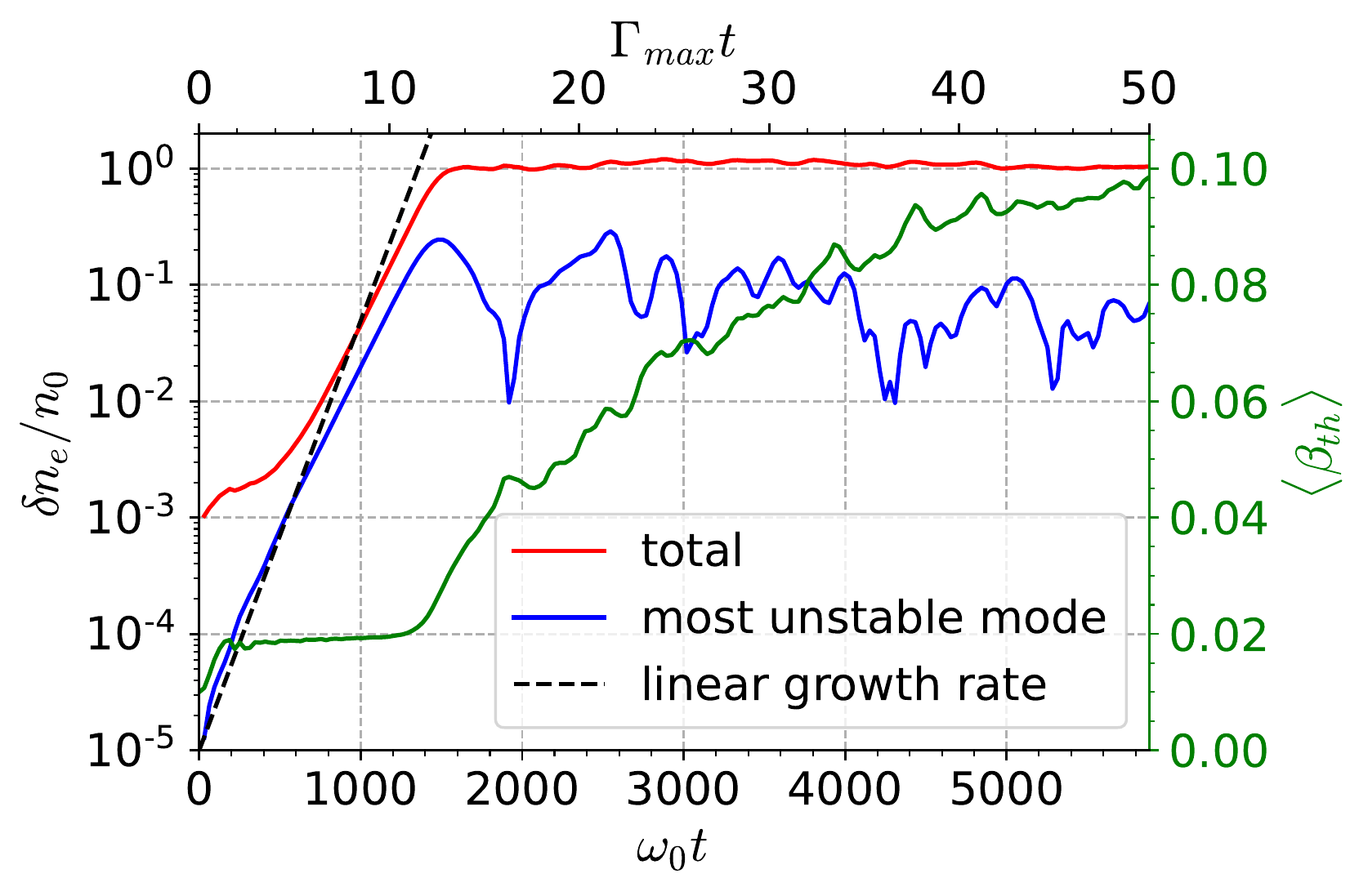}
  \includegraphics[width=8cm]{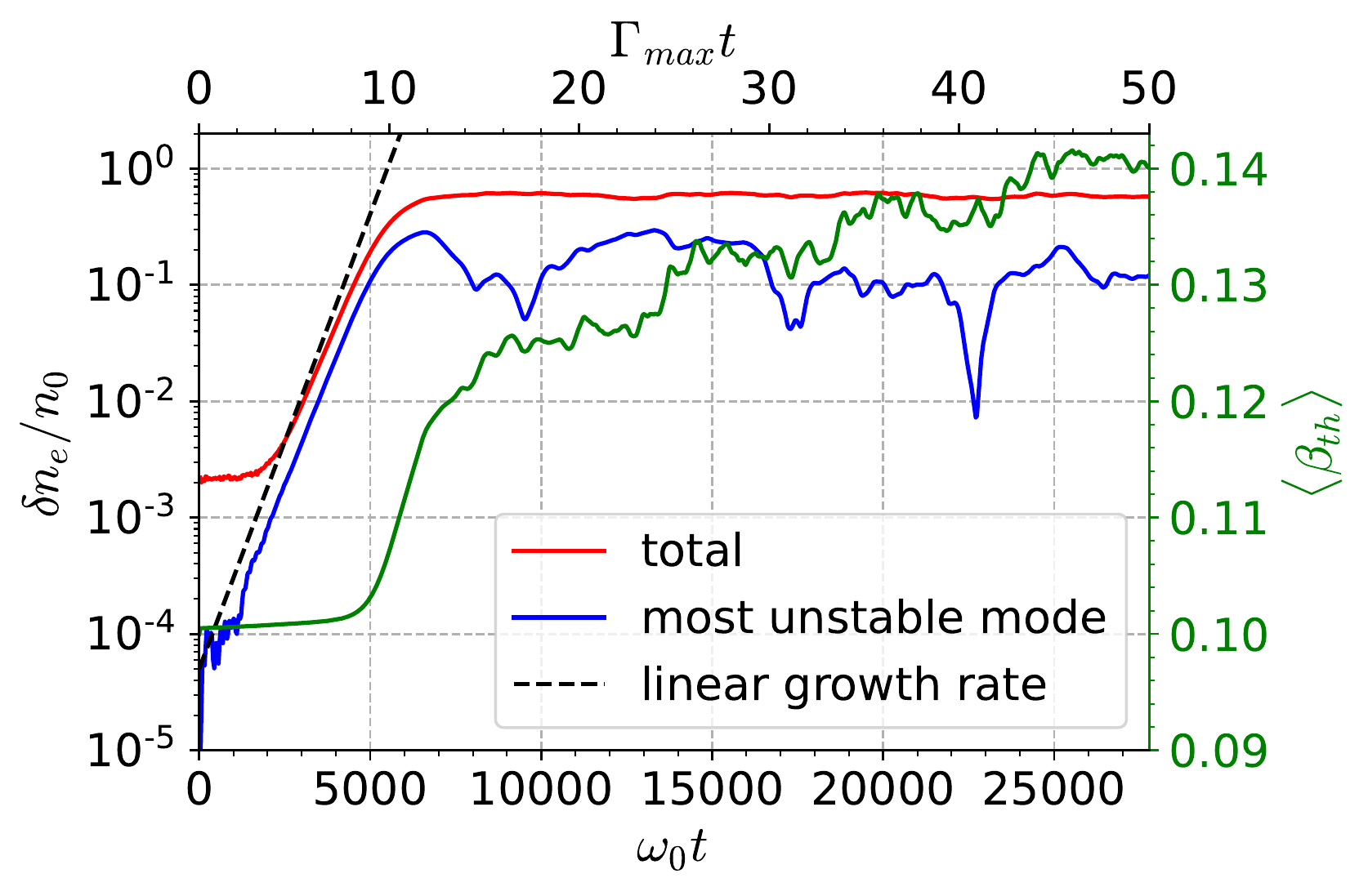}
\end{center}
\caption{Time evolution of the amplitude of the transverse electron density
fluctuations $\delta n_e$ for $\beta_{th0}=0.01$ (left) and $0.1$ (right).
The most unstable modes (blue) and total of all Fourier modes (red) are
shown. The black dashed lines represent $\propto e^{\Gamma_{max}t}$,
where $\Gamma_{max}$ is the maximum growth rate determined from linear theory
(Equation \ref{eq:gamk} for $k_x=0$). The time history of the box-averaged thermal velocity $\langle \beta_{th} \rangle$
is shown in green (axis on the right of each panel).}
\label{fig:growth}
\end{figure*}

Figure \ref{fig:growth} shows the time evolution of the transverse electron
density fluctuations $\delta n_e (y) = \sqrt{\langle n_e-n_0\rangle_x^2}$ for
$\beta_{th0}=0.01$ (left) and $0.1$ (right), where $\langle \rangle_x$
indicates the physical quantities averaged over the $x$ (pump wave
propagation) direction. 
We compute the power spectrum of $\delta n_e (y)$ and then take its square root for Figure \ref{fig:growth}.
Note that the horizontal axis range in units of $\omega_0$ is different. The most unstable modes are shown in blue.
The total of all modes (i.e., the spectrum-integrated signal), 
which is shown in red, is strongly dominated by
the most unstable mode at the linear phase 
$\Gamma_{max}t \lesssim 10$, where $\Gamma_{max}$ is the maximum growth
rate numerically determined from the linear theory (Equation \ref{eq:gamk} for $k_x=0$).
In both cases, the density filaments exponentially grow until
$\Gamma_{max}t \sim 10$ and then they get saturated.
The maximum growth rates $\Gamma_{max}$
determined from linear theory (black dashed lines)
give a good agreement with our simulation results.
 In the nonlinear phase $\Gamma_{max}t \gtrsim 10$,
the time evolution of the most unstable mode gradually deviates from the total because
the filaments begin to merge and the wavenumber of the mode with
the highest power gradually decreases, as further discussed below.

The time history of the electron thermal velocity $\langle \beta_{th} \rangle$
averaged over the whole simulation domain is shown in green (axis on the right of each panel). 
The thermal velocity is calculated in the fluid rest frame for each species.
Note that the vertical axis for $\langle \beta_{th} \rangle$ is in linear scale.
For the strong coupling regime (left in Figure \ref{fig:growth}), 
$\langle \beta_{th} \rangle$ increases for $\omega_0t \lesssim 200$ due to the SBS 
(see Appendix \ref{app:sbs}). However, most of the  heating happens during the nonlinear evolution of the FI and
we thus think that the SBS has little impact on the FI growth.
The increase of $\langle \beta_{th} \rangle$ at  early times 
is not seen for the weak coupling regime 
(right in Figure \ref{fig:growth}), demonstrating that the SBS is well-suppressed for $\beta_{th0}=0.1$.

\begin{figure*}
  \begin{center}
    \includegraphics[width=8.6cm]{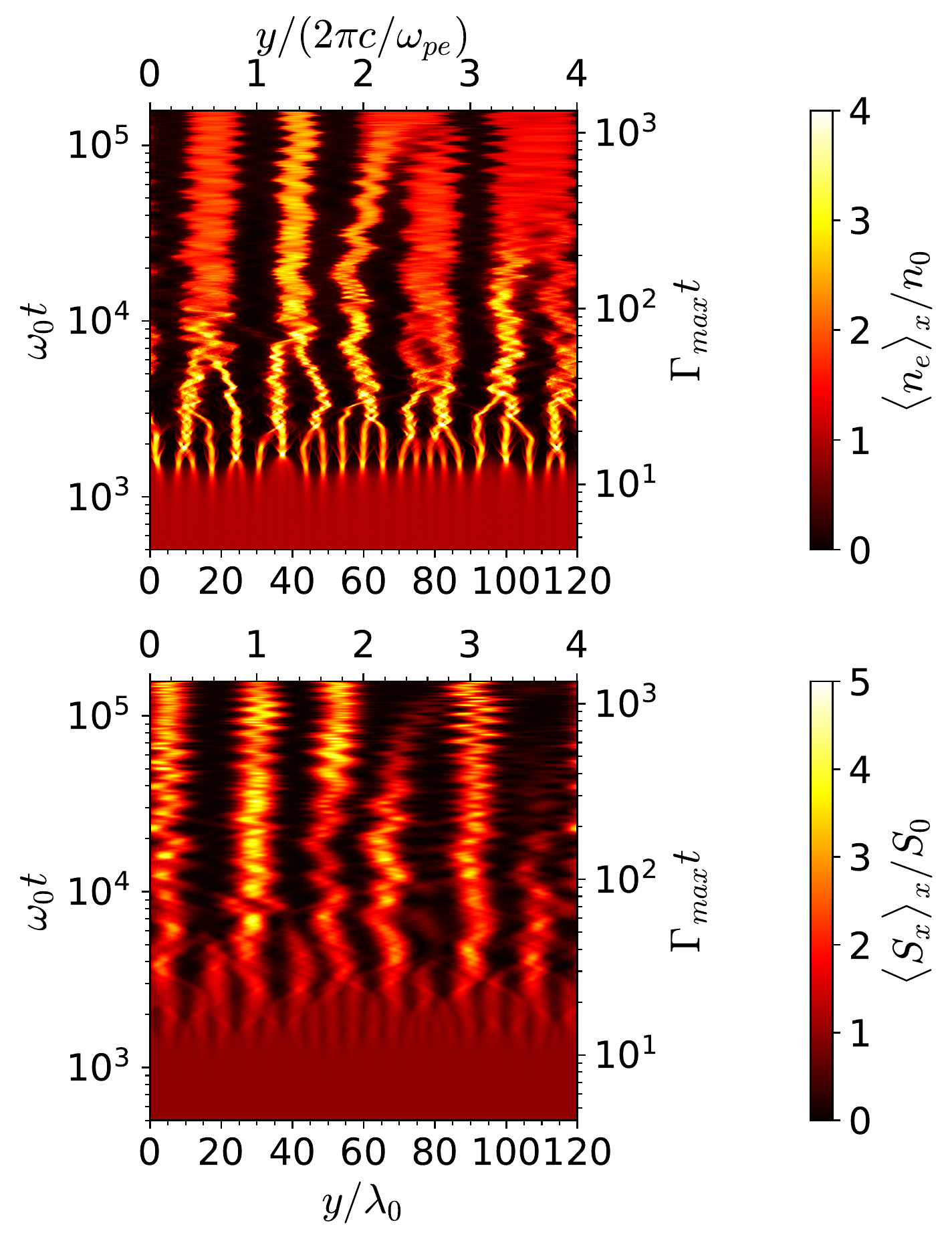}
    \includegraphics[width=8.6cm]{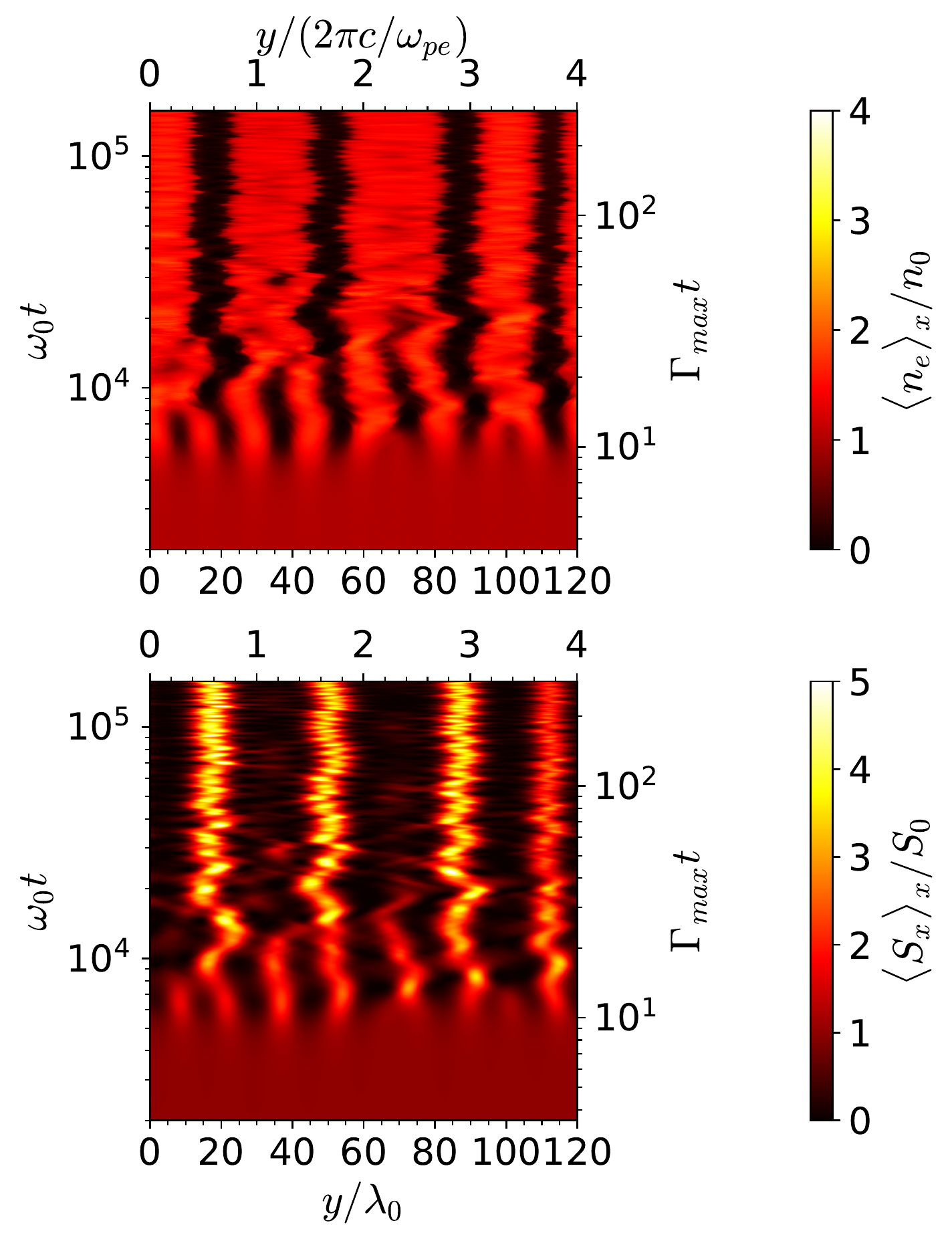}
  \end{center}
\caption{Time evolution of the $x$-averaged electron density (top panels)
and $x$ component of the $x$-averaged Poynting flux (bottom panels) for
$\beta_{th0}=0.01$ (left column) and $0.1$ (right column).}
\label{fig:evo}
\end{figure*}

Figure \ref{fig:evo} shows the temporal evolution of the $x$-averaged
electron density $\langle n_e \rangle_x$ (top panels) and $x$ component of the
Poynting flux $\langle S_x \rangle_x$ (bottom panels) for $\beta_{th0}=0.01$
(left column) and $0.1$ (right column), where $\langle S_x \rangle_x$ is
normalized by the initial mean flux $S_0=E_0^2/8\pi$.
In the linear phase $\Gamma_{max}t \lesssim 10$,
the amplitude of the density filaments for
$\beta_{th0} = 0.01$
is larger than for $\beta_{th0} = 0.1$,
because colder plasmas are more easily compressed by the wave ponderomotive force
due to their weaker pressure gradients.
In the final state of our simulations, the density amplitudes are comparable between the two cases,
because the plasma gets heated during the nonlinear evolution of the FI 
and the temperatures become comparable in the two cases, as shown with the grey lines in 
Figure \ref{fig:growth} and further discussed in Section \ref{sec:sat}.
The density filaments gradually merge for $\Gamma_{max}t \gtrsim 10$
and the filament merging continues until the wavelength of the filament
reaches  $\sim 2 \pi c/\omega_{pe}$, i.e., comparable to the electron skin depth.
We discuss the saturation of the filament merging in Section \ref{sec:sat}.
The wave Poynting flux peaks in the lower
density regions, i.e., the wave power accumulates in the density
cavities. The electromagnetic waves then propagate between the density filaments
as in a waveguide.

\begin{figure*}
\begin{center}
  \includegraphics[width=8.6cm]{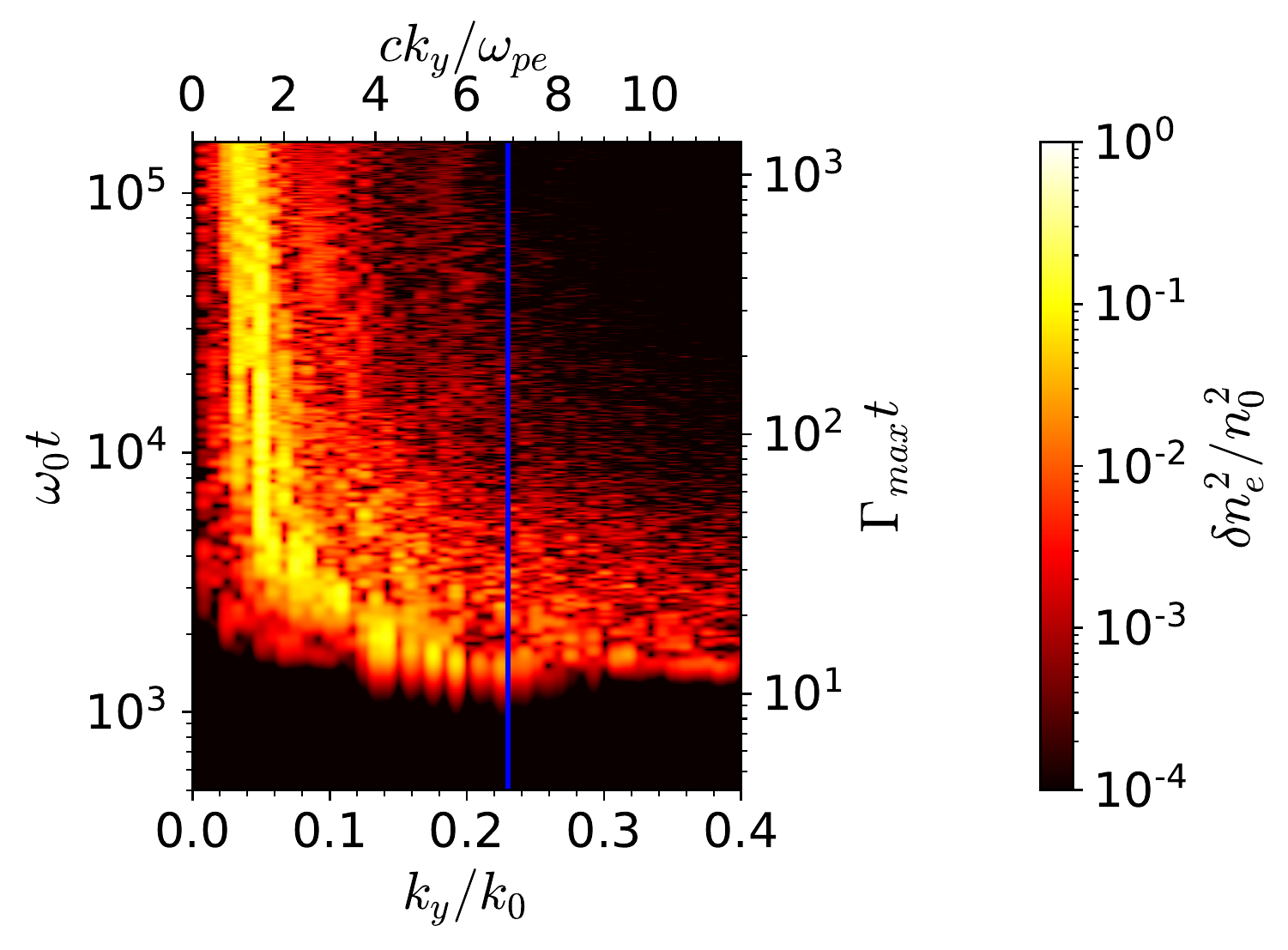}
  \includegraphics[width=8.6cm]{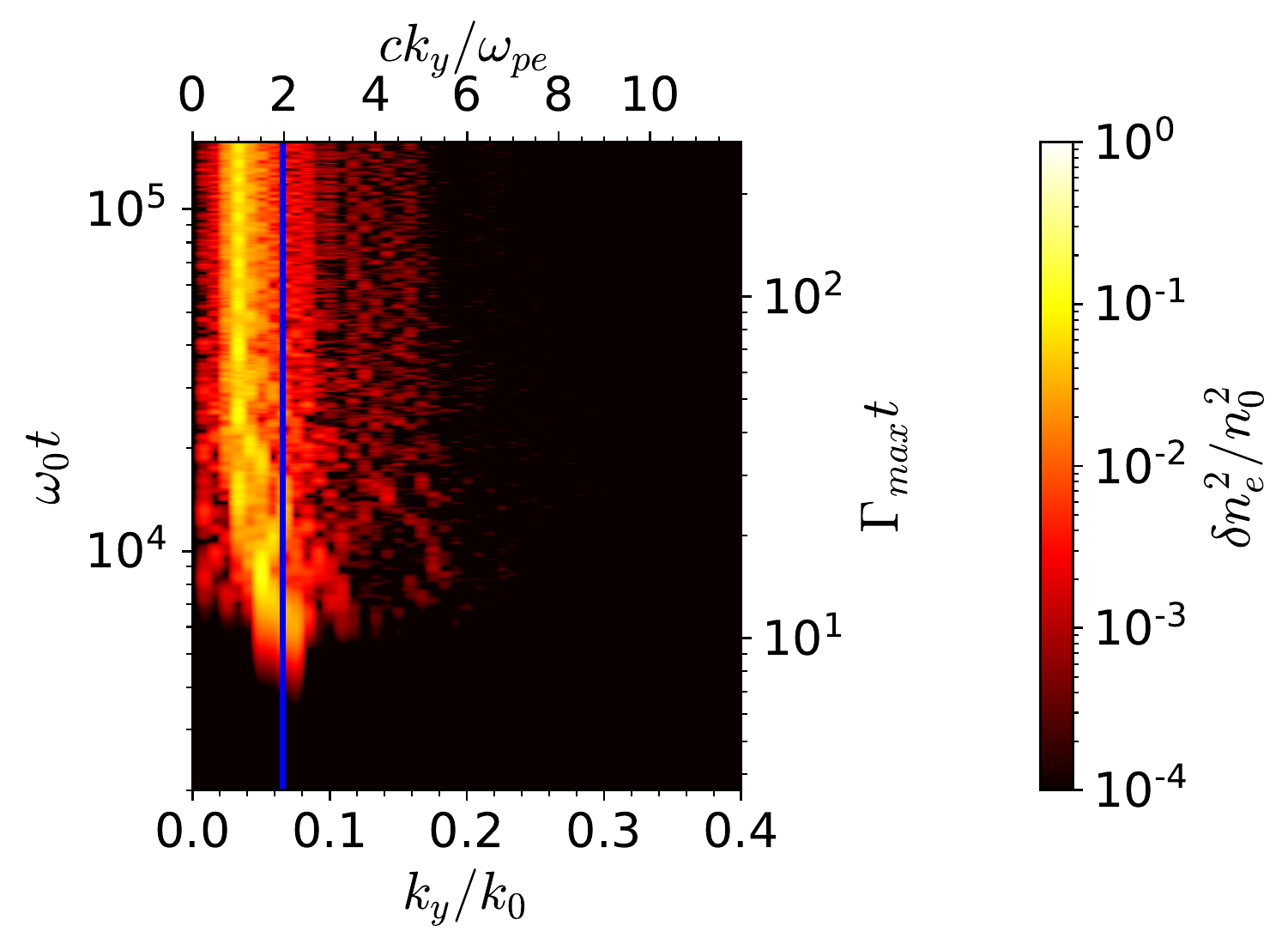}
\end{center}
\caption{Time evolution of the power spectra of the $x$-averaged electron density
fluctuations for $\beta_{th0}=0.01$ (left) and $0.1$ (right).
The blue lines indicate the theoretical fastest-growing modes.}
\label{fig:power}
\end{figure*}

Figure \ref{fig:power} shows the time evolution of the power spectra
of the $x$-averaged electron density fluctuations
for $\beta_{th0}=0.01$ (left) and $0.1$ (right).
The blue lines correspond to the wavenumber of the theoretical fastest-growing
modes: $k_y/k_0 \sim 0.2$ for $\beta_{th0}=0.01$ and
$k_y/k_0 \sim 0.07$ for $\beta_{th0}=0.1$ in Figure \ref{fig:linear}.
The observed peaks at the linear stage $\Gamma_{max}t \lesssim 10$ are
consistent with the theoretical estimates.
The most unstable wavenumber gradually decreases down to $\sim \omega_{pe}/c$.

\subsection{Saturation Mechanism of Filament Merging}\label{sec:sat}

\begin{figure*}
\begin{center}
  \includegraphics[width=8.6cm]{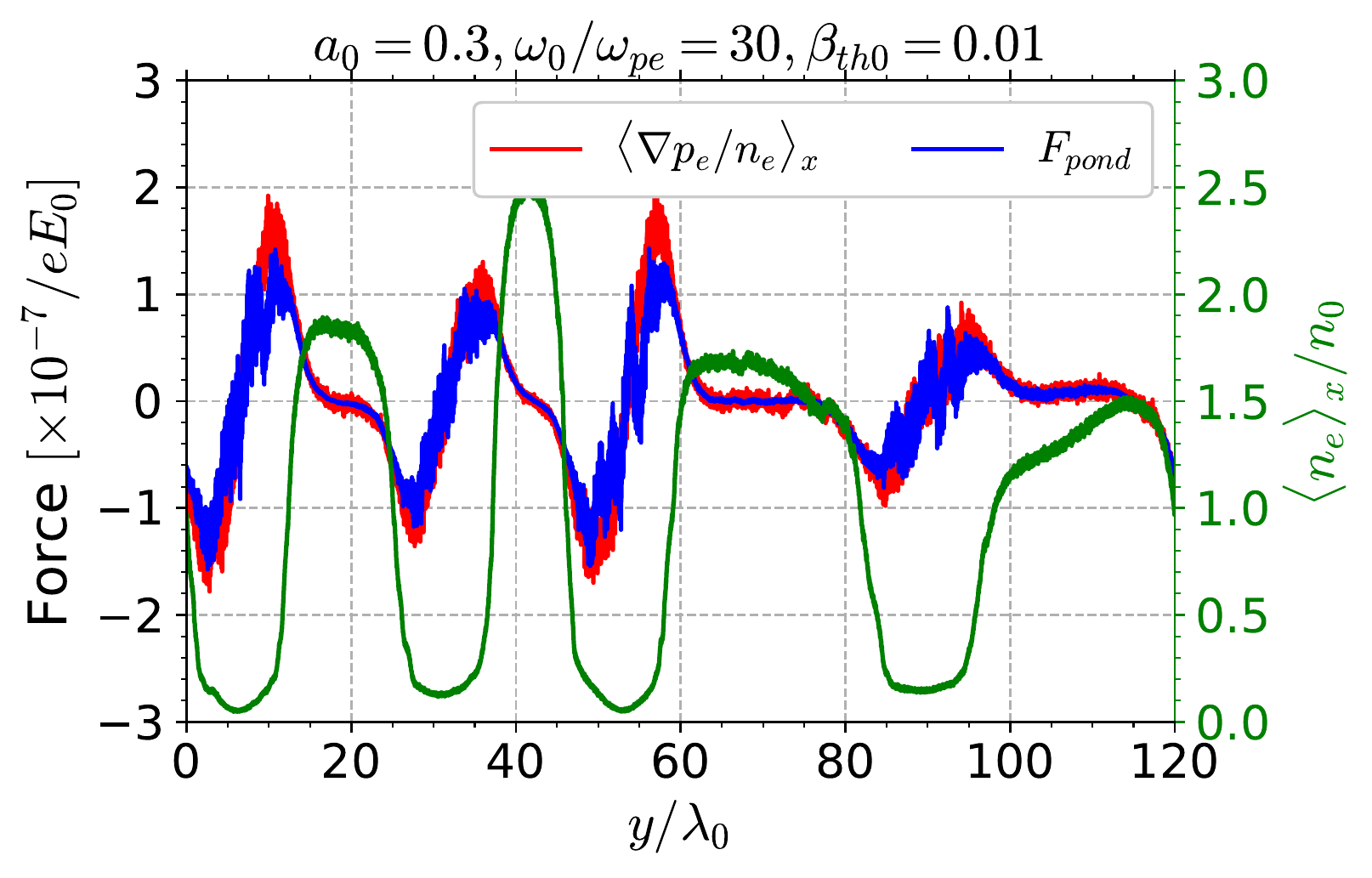}
  \includegraphics[width=8.6cm]{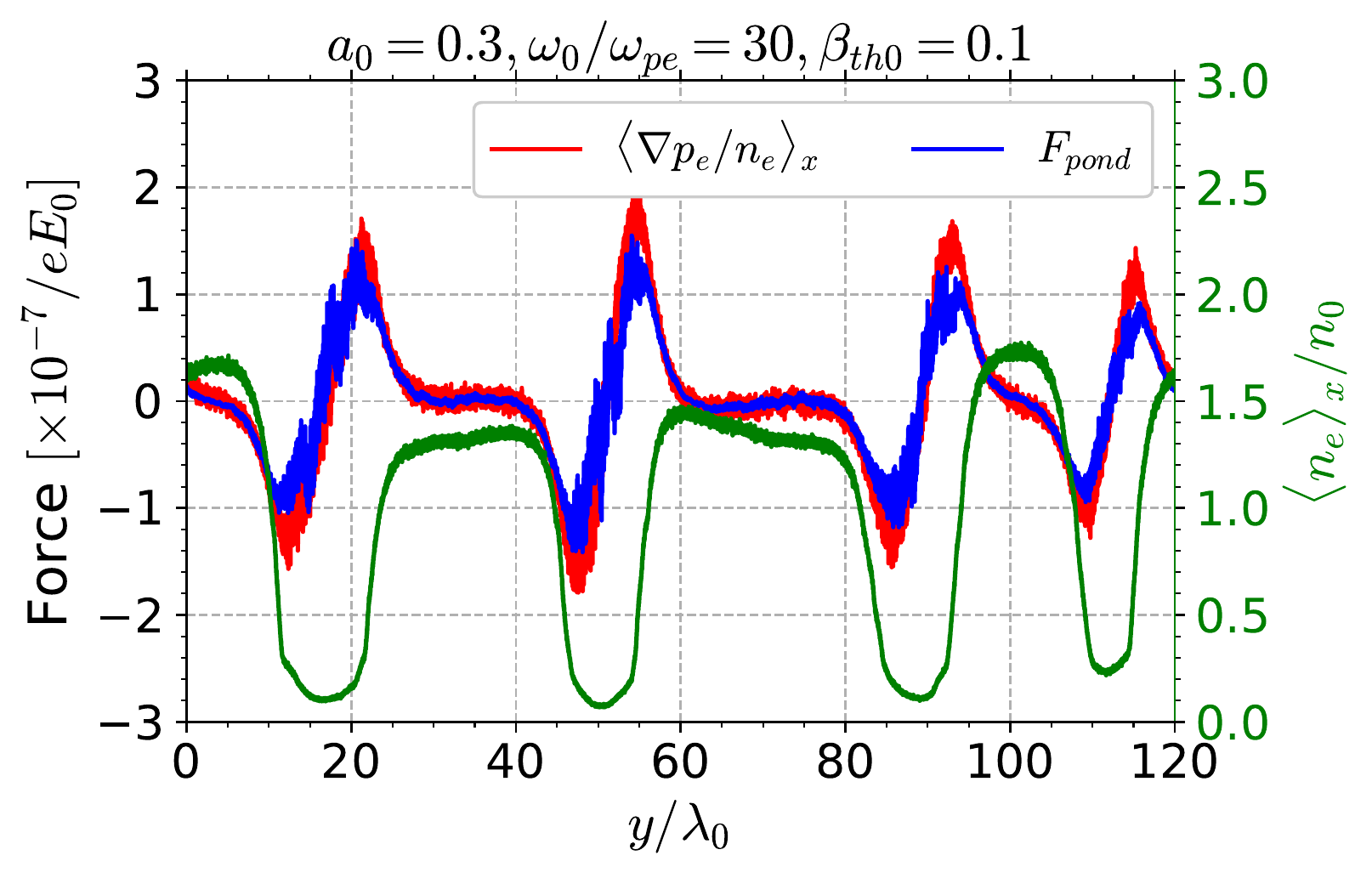}
\end{center}
\caption{Snapshots of the ponderomotive force (blue) and
pressure gradient (red) at the final state $\omega_0t=157254$ for
$\beta_{th0}=0.01$ (left) and $0.1$ (right).
The green lines indicate the $x$-averaged electron density.}
\label{fig:force}
\end{figure*}

The FI saturates when force balance between
the pressure gradient and ponderomotive force is achieved
\citep{Kaw1973,Sobacchi2022b}.
The ponderomotive force exerted by the electromagnetic wave
expels particles from the region of high intensity.
The pressure gradient is gradually amplified by the
compression and it finally balances the ponderomotive force.
Figure \ref{fig:force} shows
snapshots of the $x$-averaged ponderomotive force (blue) and plasma
pressure (red) normalized by $eE_0$
at the final state of our simulations $\omega_0t=157254$ for 
$\beta_{th0}=0.01$ (left) and $0.1$ (right).
The pressure gradient $\nabla p_e$ is the $y$ derivative of the
$yy$ component of the pressure tensor and averaged over the $x$ direction.
The ponderomotive force is by definition the sum of the advection and
nonlinear Lorentz force averaged over the wave period.
We determine the $y$ component of the ponderomotive force $F_{pond}$
for electrons from the snapshots averaged over the $x$ direction
(i.e., one wavelength of the pump wave),
\begin{equation}
  F_{pond}
  = \left \langle - (\bvec{v_e} \cdot \bvec{\nabla})  v_{ey}
  -\frac{e}{m_ec} (v_{ez}B_x-v_{ex}B_z) \right\rangle_x
\end{equation}
The green lines indicate the $x$-averaged electron density.
The electromagnetic waves escape from the higher density region 
as shown in the bottom panels of Figure \ref{fig:evo},
and thus the ponderomotive force vanishes there.
The force balance between the pressure gradient and ponderomotive force
is clearly achieved across the whole transverse direction.

\begin{figure*}
\begin{center}
  \includegraphics[width=8.6cm]{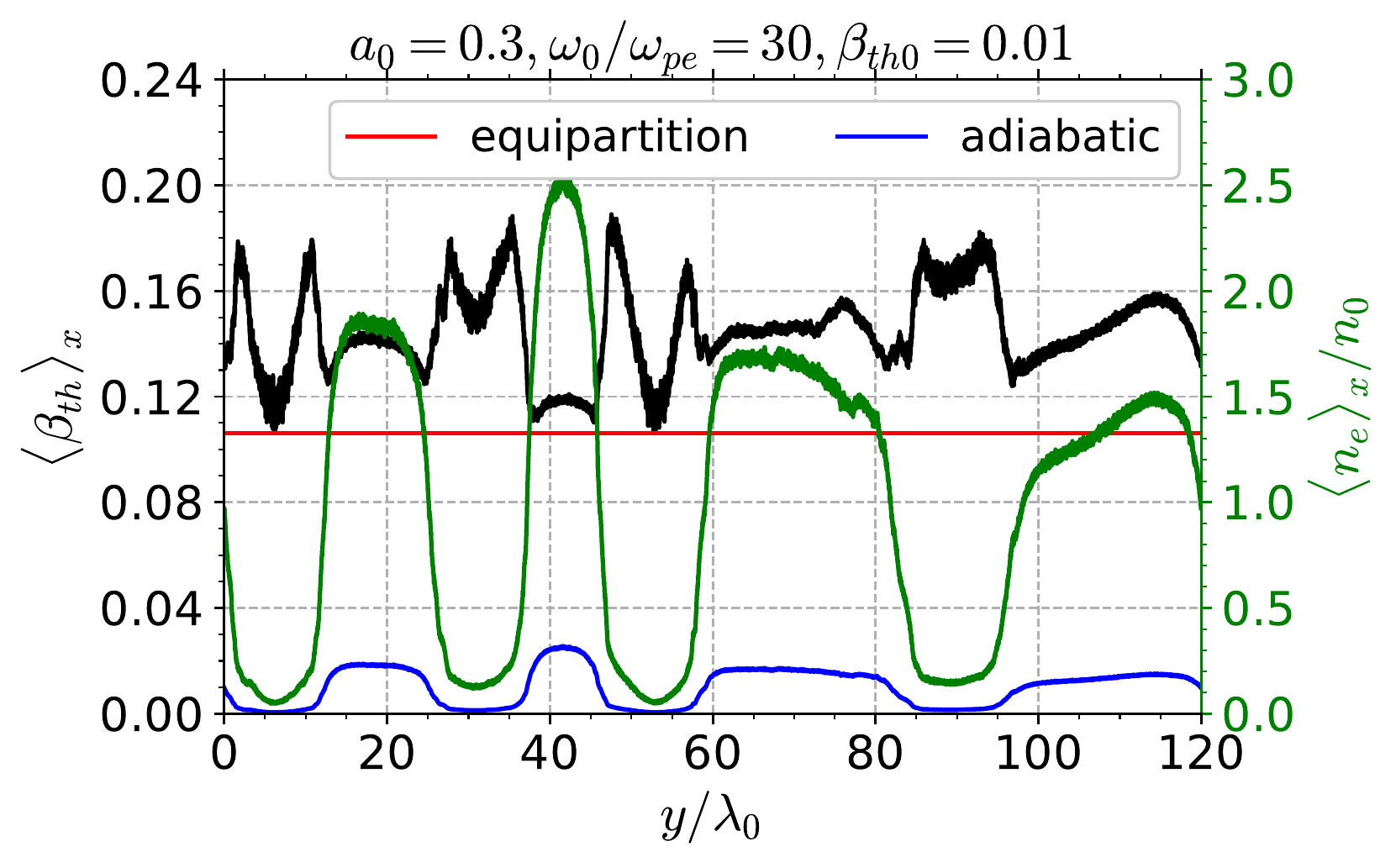}
  \includegraphics[width=8.6cm]{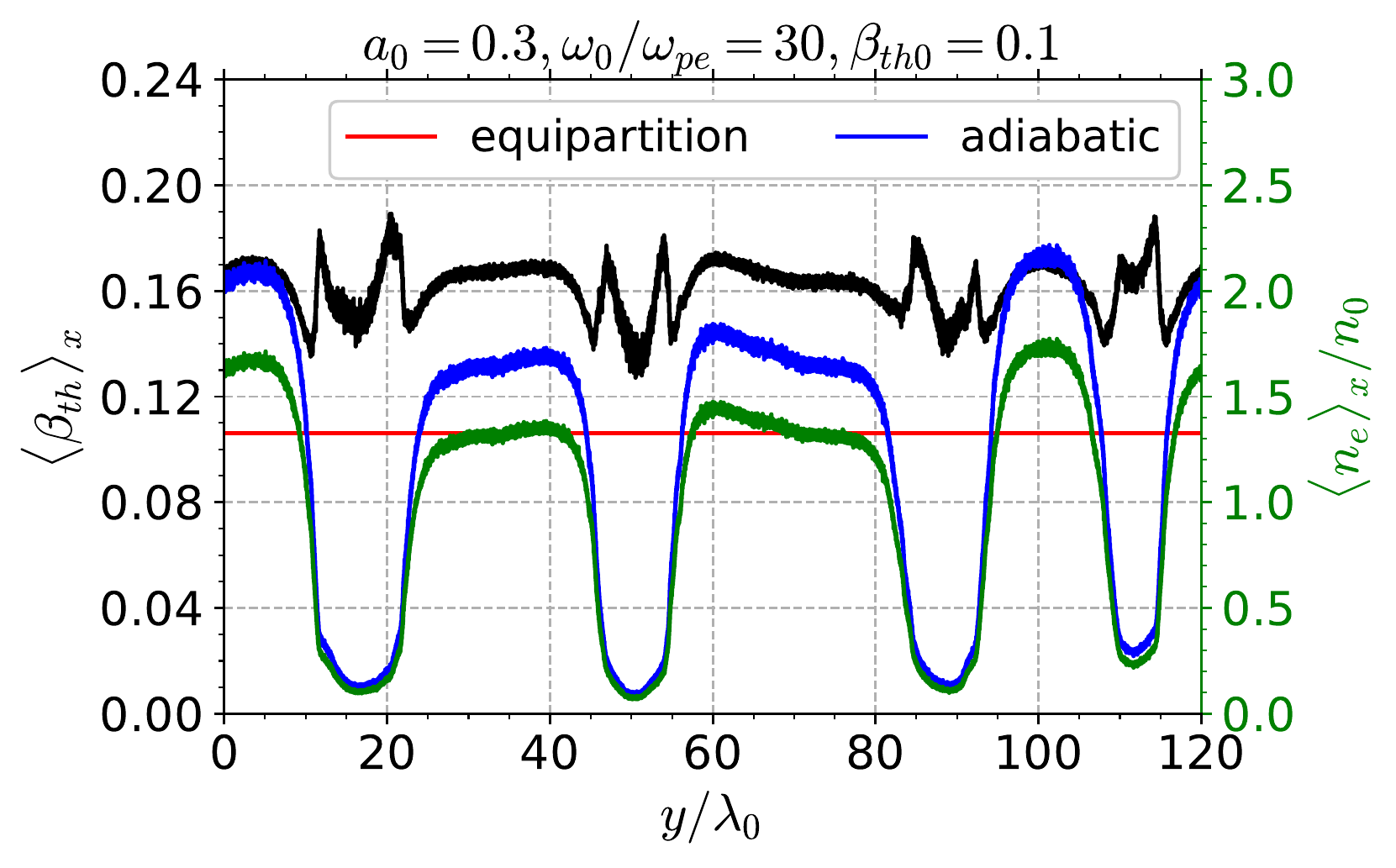}
\end{center}
\caption{Snapshots of the electron thermal velocity (black)
at the final state $\omega_0t=157254$ 
for $\beta_{th0}=0.01$ (left) and $0.1$ (right).
The $x$-averaged electron density is shown in green.
The thermal velocity determined from the 
adiabatic relation $\beta_{th}^2/n_e^2=const.$ is shown in blue. The red lines 
indicate the thermal velocity for the energy equipartition.}
\label{fig:vte}
\end{figure*}

\cite{Sobacchi2022b} discussed the saturation mechanism of the FI based
on the assumption that the adiabatically-compressed density filaments are
supported by the ponderomotive force in the steady state. They pointed out 
that non-adiabatic heating can be important for the strong coupling regime 
and it can raise the plasma temperature
because the force balance between the ponderomotive force and the pressure
gradient does not have time to be established for
$\tau_{grow} \ll \tau_{cross}$, where $\tau_{grow}$ is the e-folding time of the 
FI and $\tau_{cross}$ is the sound crossing time of the density filaments.
To investigate the effect of the non-adiabatic heating, we measure the thermal velocity 
in our simulations. 
Figure \ref{fig:vte} shows the electron thermal velocity (black)
at the final state of our simulations $\omega_0t=157254$, which is the 
same time as Figure \ref{fig:force},
for $\beta_{th0}=0.01$ (left) and $0.1$ (right).
The green lines indicate the $x$-averaged electron density.
If only the adiabatic compression contributes to the plasma heating,
the thermal velocity satisfies
\begin{equation}
  \frac{\beta_{th}^2}{n_e^{\gamma_{ad}-1}}=const.,
\end{equation}
The adiabatic thermal velocity is determined from the measured
density profile adopting a choice of $\gamma_{ad}=3$ and shown in blue.
For the weak coupling case $\beta_{th0} = 0.1$ (right in Figure \ref{fig:vte}),
the thermal velocity in the higher density region is well-explained by the
adiabatic heating. The non-adiabatic heating operates in the density
cavity and is associated with the filament mergers.
For the strong coupling regime $\beta_{th0}=0.01$ (left in Figure \ref{fig:vte}), 
the thermal velocity at the final time is much larger than the adiabatic heating, indicating that 
the non-adiabatic heating is dominant.

The non-adiabatic heating may saturate when the
equipartition between the ponderomotive potential and total (electron +
positron) thermal energy is achieved.
Since the initial ponderomotive potential is $m_ec^2a_0^2/4$,
the equipartition thermal energy is $m_ec^2a_0^2/8$ and
the thermal velocity is thus
\begin{equation}
  \beta_{th} \sim \frac{a_0}{2\sqrt{2}},
\end{equation}
which is shown in red in Figure \ref{fig:vte}.
This estimate is roughly consistent with the measured thermal velocity
at the final time. The filament merging continues until the wavelength of the 
filament reaches $\sim 2 \pi c/\omega_{pe}$
as already shown. The saturation wavelength may be explained by an argument 
relying as well on the saturation thermal velocity.
If the linear analysis is still valid at the saturation stage,
Equation \ref{eq:kyw} for $\beta_s \sim a_0/2\sqrt{2}$
reduces to $k_y \sim \omega_{pe}/c$  in the weak coupling case. 
In the strong coupling case, Equation \ref{eq:kys} reduces to
\begin{equation}
  \frac{\omega_{pe}}{c} < k_y < \frac{2\sqrt{2}\omega_{pe}}{c},
\end{equation}
where $\omega_0 \gg \omega_{pe}/a_0$ is applied.
The wavenumber of the most unstable mode may gradually approach
the inverse skin depth due to the non-adiabatic heating.

\section{Summary and Discussion}\label{sec:sum}

We study the nonlinear evolution of the filamentation instability (FI) of strong electromagnetic waves in pair plasmas using 2D PIC simulations. Our simulations show that the FI generates transverse density filaments 
and that the electromagnetic waves propagate in near vacuum between the density filaments, as in a waveguide. 
We find that the density filaments merge until the filament wavelength reaches the electron skin depth. The filament merging ceases when force balance between the ponderomotive force and the pressure gradient is established. Non-adiabatic heating operates during the evolution of the FI and can be important especially in the strong coupling regime, i.e. when the e-folding time of the FI is shorter than the sound travel time across the filaments. Non-adiabatic heating may saturate when equipartition between the ponderomotive potential and the plasma thermal energy is achieved.

\begin{figure}
\begin{center}
  \includegraphics[width=8cm]{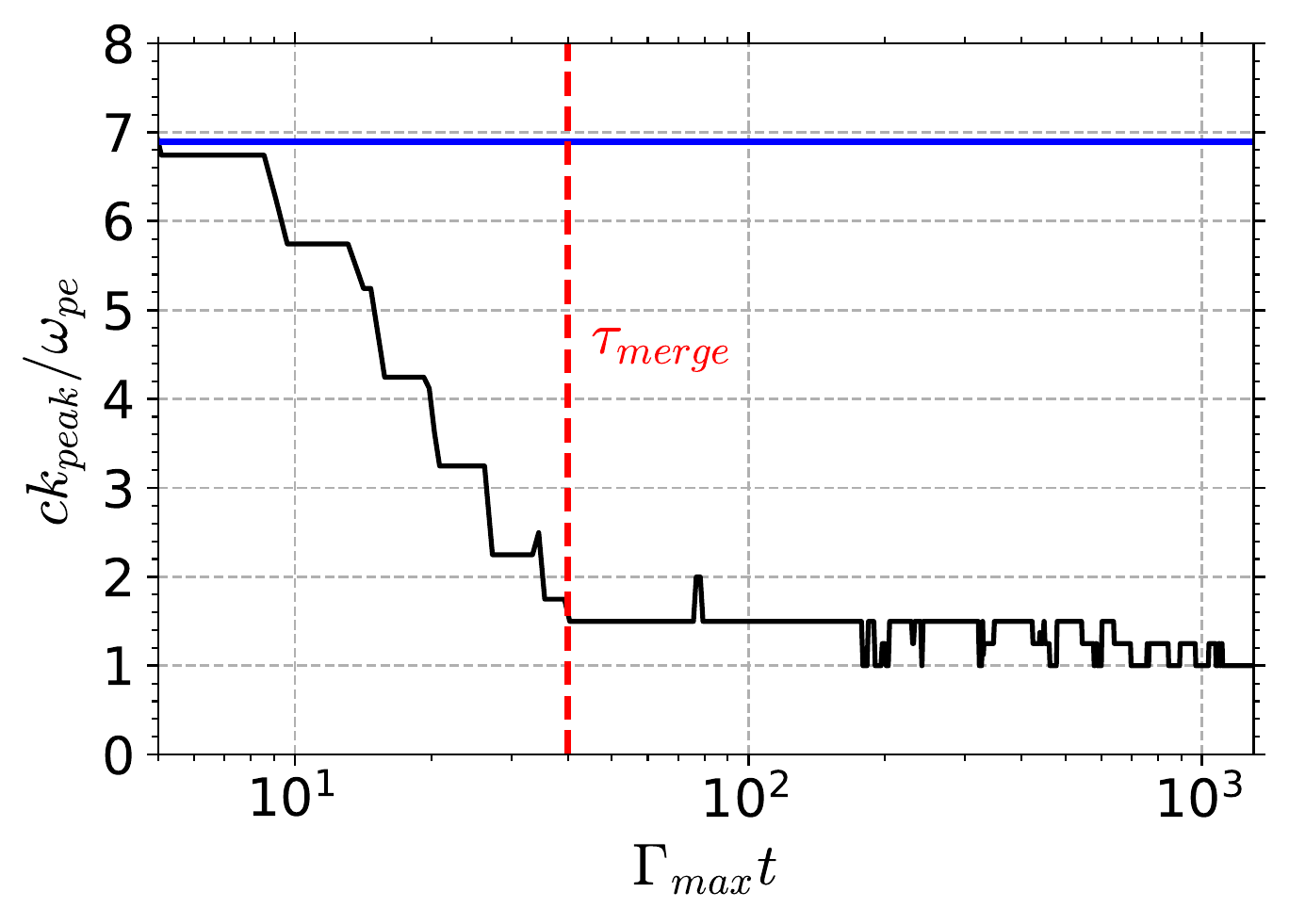}
\end{center}
\caption{Time evolution of the peak wavenumber for the strong coupling regime 
$\beta_{th0}=0.01$ in the left panel of
Figure \ref{fig:power}. The blue solid line indicates the fastest-growing
mode from the linear theory.}
\label{fig:kpeak}
\end{figure}

We now discuss the implications of our results for Fast Radio Bursts (FRBs). 
The FRB propagation has four important time scales: (i) the time scale on which the FI exponentially grows, $\tau_{FI}$, (ii) the filament merging time scale $\tau_{merge}$, (iii) the pulse duration time $\tau_{pulse}$, and (iv) the expansion time of the wave front $\tau_{exp}$. We estimate $\tau_{merge}$ from our simulations in the strong coupling
regime, as appropriate for FRBs \citep{Sobacchi2022b}.
Figure \ref{fig:kpeak} shows the time evolution of the peak wavenumber
of the power spectrum (taken from the left panel of Figure \ref{fig:power}).
The blue solid line indicates the fastest-growing modes from linear
theory, which agrees with the simulation results in the linear phase
$t \lesssim \tau_{FI}\sim 10/\Gamma_{max}$. Since the peak wavenumber exponentially
decreases until $\Gamma_{max}t \sim 40$,
we define the merging time as $\tau_{merge} \sim 4\tau_{FI}\sim 40/\Gamma_{max}$.
Evaluating $\Gamma_{max}$ from  linear theory, the merging time $\tau_{merge}$ in the rest frame of the magnetar wind can then be estimated as
\begin{multline}
  \tau_{merge} \sim 80\, {\rm ms}
  \left(\frac{L}{10^{42} {\rm\; erg \ s^{-1}}}\right)^{-\frac{1}{2}}
  \left(\frac{\dot{N}}{10^{39} {\rm\; s^{-1}}}\right)^{-\frac{1}{2}} \\
  \times \left(\frac{\gamma_w}{10^2}\right)^{\frac{1}{2}}
  \left(\frac{\nu_{obs}}{1 {\rm\; GHz}} \right)
  \left(\frac{R}{10^{14} {\rm\; cm}}\right)^2,
\end{multline}
where $L$ is the observed radio luminosity, $\dot{N}$ is the particle outflow
rate, $\gamma_{w}$ is the wind bulk Lorentz factor, $\nu_{obs}$ is the obserbed
radio frequency, and $R$ is the distance from the source
\citep{Beloborodov2020,Sobacchi2022b}.
The time duration of the radio pulse in the wind rest frame $\tau_{pulse}$ is
\begin{equation}
  \tau_{pulse} = 2\gamma_{w}\tau_{obs}
  \sim 200  \,{\rm ms}\left(\frac{\gamma_w}{10^2}\right)\left(\frac{\tau_{obs}}{1 {\rm\; ms}}\right),
\end{equation}
where $\tau_{obs}$ is the observed pulse duration.
The expansion time of the wave front in the wind frame is
\begin{equation}
  \tau_{exp} = \frac{R}{2\gamma_wc}
  \sim 20 \,{\rm s}\left(\frac{R}{10^{14} {\rm\; cm}}\right)\left(\frac{\gamma_w}{10^2}\right)^{-1}.
\end{equation}
Since $\tau_{FI}\lesssim\tau_{merge}  \lesssim \tau_{pulse} \ll \tau_{exp}$, the radio wave is filamented, and the filaments merge before the radio pulse can propagate through the unperturbed plasma ahead of the wave front.

The merging time may get longer for the realistic case in which the peak wavenumber in the 
linear stage is $\gg \omega_{pe}/c$, a case we cannot achieve due to computational 
limitations. Then the filamentation instability may develop in the regime where the merging 
time is longer than the duration of the radio pulse, i.e. 
$\tau_{FI} \lesssim \tau_{pulse}\lesssim \tau_{merge} \ll \tau_{exp}$. 
In this regime the evolution of the filaments on the time scale $\tau_{merge}$ is unclear. Our 
simulations  employ a  periodic boundary condition in the wave propagation direction, so the 
radio pulse continuously interacts with density filaments. In contrast, for realistic FRB 
conditions the density filaments are non-propagating and stop interacting with the radio pulse 
after the time scale $\tau_{pulse}$. Then the pulse should propagate through an unperturbed 
plasma ahead of the wave front. We will study the effect of more realistic boundary 
conditions---including a self-consistent description of wave propagation---in a future 
publication.

We assumed that the initial velocity distribution is isotropic in this study.
When plasmas are highly magnetized, which is the case for the magnetar wind, 
a temperature anisotropy is generally expected.
\cite{Sobacchi2022a} discussed the effect of the ambient magnetic field and 
demonstrated that the FI is independent on the thermal velocity 
in the direction perpendicular to the ambient magnetic field because the ponderomotive force 
preferentially pushes particles along the parallel direction. 
Therefore, we expect that the FI should be primarily affected by the parallel temperature 
for the anisotropic velocity distribution. 

Our results are valid for the weak pump wave condition $a_0 \ll 1$, in which
the particle oscillation velocities in the wave fields are
much smaller than the speed of light.
The radio pulses are much stronger near the FRB progenitors
and $a_0 \gtrsim 1$ can be satisfied for $R \lesssim 10^{13} {\rm cm}$
\citep{Luan2014,Beloborodov2020}. In this relativistic regime,  higher order couplings
$(\omega_n, \bvec{k_n}) = (\omega +n\omega_0, \bvec{k}+ n\bvec{k_0})$,
where $n = \pm 2,\pm 3,\pm 4,...$ is an integer, are no longer negligible.
We will explore the relativistic regime $a_0 \gtrsim 1$ in a future publication.

\section*{Acknowledgements}

MI is grateful to Richard Sydora and Shuichi Matsukiyo 
for fruitful discussions.
MI acknowledges support from JSPS KAKENHI grant No. 20J00280 and 20KK0064.
LS acknowledges support from NASA80NSSC18K1104. 
This research was facilitated by 
Multimessenger Plasma Physics Center (MPPC), NSF grant PHY-2206607. 
This work used the computational resources of the HPCI system provided by 
Information Technology Center, Nagoya University through the HPCI System Research Project (Project ID: hp220041). 
Numerical computations were in part carried out on Cray
XC50 at Center for Computational Astrophysics, National Astronomical observatory 
of Japan.

\section*{DATA AVAILABILITY}

The simulation code is available on request.

\bibliographystyle{mnras}
\bibliography{ref}

\begin{thebibliography}{}
\makeatletter
\relax
\def\mn@urlcharsother{\let\do\@makeother \do\$\do\&\do\#\do\^\do\_\do\%\do\~}
\def\mn@doi{\begingroup\mn@urlcharsother \@ifnextchar [ {\mn@doi@}
  {\mn@doi@[]}}
\def\mn@doi@[#1]#2{\def\@tempa{#1}\ifx\@tempa\@empty \href
  {http://dx.doi.org/#2} {doi:#2}\else \href {http://dx.doi.org/#2} {#1}\fi
  \endgroup}
\def\mn@eprint#1#2{\mn@eprint@#1:#2::\@nil}
\def\mn@eprint@arXiv#1{\href {http://arxiv.org/abs/#1} {{\tt arXiv:#1}}}
\def\mn@eprint@dblp#1{\href {http://dblp.uni-trier.de/rec/bibtex/#1.xml}
  {dblp:#1}}
\def\mn@eprint@#1:#2:#3:#4\@nil{\def\@tempa {#1}\def\@tempb {#2}\def\@tempc
  {#3}\ifx \@tempc \@empty \let \@tempc \@tempb \let \@tempb \@tempa \fi \ifx
  \@tempb \@empty \def\@tempb {arXiv}\fi \@ifundefined
  {mn@eprint@\@tempb}{\@tempb:\@tempc}{\expandafter \expandafter \csname
  mn@eprint@\@tempb\endcsname \expandafter{\@tempc}}}

\bibitem[\protect\citeauthoryear{Andersen et~al.,}{Andersen
  et~al.}{2020}]{Andersen2020}
Andersen B.~C.,  et~al., 2020, \mn@doi [Nature] {10.1038/s41586-020-2863-y},
  587, 54

\bibitem[\protect\citeauthoryear{Babul \& Sironi}{Babul \&
  Sironi}{2020}]{Babul2020}
Babul A.-N.,  Sironi L.,  2020, \mn@doi [\mnras] {10.1093/mnras/staa2612}, 499,
  2884

\bibitem[\protect\citeauthoryear{Beloborodov}{Beloborodov}{2017}]{Beloborodov2017}
Beloborodov A.~M.,  2017, \mn@doi [\apjl] {10.3847/2041-8213/aa78f3}, 843, L26

\bibitem[\protect\citeauthoryear{Beloborodov}{Beloborodov}{2020}]{Beloborodov2020}
Beloborodov A.~M.,  2020, \mn@doi [\apj] {10.3847/1538-4357/ab83eb}, 896, 142

\bibitem[\protect\citeauthoryear{Bochenek, Ravi, Belov, Hallinan, Kocz,
  Kulkarni  \& McKenna}{Bochenek et~al.}{2020}]{Bochenek2020}
Bochenek C.~D.,  Ravi V.,  Belov K.~V.,  Hallinan G.,  Kocz J.,  Kulkarni
  S.~R.,   McKenna D.~L.,  2020, \mn@doi [Nature] {10.1038/s41586-020-2872-x},
  587, 59

\bibitem[\protect\citeauthoryear{Cohen \& Max}{Cohen \& Max}{1979}]{Cohen1979}
Cohen B.~I.,  Max C.~E.,  1979, \mn@doi [Phys. Fluids] {10.1063/1.862713}, 22,
  1115

\bibitem[\protect\citeauthoryear{Cohen, Divol, Langdon  \& Williams}{Cohen
  et~al.}{2005}]{Cohen2005}
Cohen B.~I.,  Divol L.,  Langdon A.~B.,   Williams E.~A.,  2005, \mn@doi [Phys.
  Plasmas] {10.1063/1.1878792}, 12, 1

\bibitem[\protect\citeauthoryear{Day et~al.,}{Day et~al.}{2020}]{Day2020}
Day C.~K.,  et~al., 2020, \mn@doi [\mnras] {10.1093/mnras/staa2138}, 497, 3335

\bibitem[\protect\citeauthoryear{Deutsch, Furukawa, Mima, Murakami  \&
  Nishihara}{Deutsch et~al.}{1996}]{Deutsch1996}
Deutsch C.,  Furukawa H.,  Mima K.,  Murakami M.,   Nishihara K.,  1996,
  \mn@doi [\prl] {10.1103/PhysRevLett.77.2483}, 77, 2483

\bibitem[\protect\citeauthoryear{Drake, Kaw, Lee, Schmid, Liu  \&
  Rosenbluth}{Drake et~al.}{1974}]{Drake1974}
Drake J.~F.,  Kaw P.~K.,  Lee Y.~C.,  Schmid G.,  Liu C.~S.,   Rosenbluth
  M.~N.,  1974, \mn@doi [Phys. Fluids] {10.1063/1.1694789}, 14, 778

\bibitem[\protect\citeauthoryear{Edwards, Fisch  \& Mikhailova}{Edwards
  et~al.}{2016}]{Edwards2016}
Edwards M.~R.,  Fisch N.~J.,   Mikhailova J.~M.,  2016, \mn@doi [\prl]
  {10.1103/PhysRevLett.116.015004}, 116, 015004

\bibitem[\protect\citeauthoryear{Esirkepov}{Esirkepov}{2001}]{Esirkepov2001}
Esirkepov T.,  2001, \mn@doi [Comput. Phys. Commun.]
  {10.1016/S0010-4655(00)00228-9}, 135, 144

\bibitem[\protect\citeauthoryear{Forslund, Kindel  \& Lindman}{Forslund
  et~al.}{1975}]{Forslund1975}
Forslund D.~W.,  Kindel J.~M.,   Lindman E.~L.,  1975, \mn@doi [Phys. Fluids]
  {10.1063/1.861248}, 18, 1002

\bibitem[\protect\citeauthoryear{Fried \& Conte}{Fried \&
  Conte}{1961}]{Fried1961}
Fried B.~D.,  Conte S.~D.,  1961, The Plasma Dispersion Function: The Hilbert
  Transform of the Gaussian.
Academic Press, New York

\bibitem[\protect\citeauthoryear{Ghosh, Kagan, Keshet  \& Lyubarsky}{Ghosh
  et~al.}{2022}]{Ghosh2022}
Ghosh A.,  Kagan D.,  Keshet U.,   Lyubarsky Y.,  2022, \mn@doi [\apj]
  {10.3847/1538-4357/ac581d}, 930, 106

\bibitem[\protect\citeauthoryear{Ikeya \& Matsumoto}{Ikeya \&
  Matsumoto}{2015}]{Ikeya2015}
Ikeya N.,  Matsumoto Y.,  2015, \mn@doi [\pasj] {10.1093/pasj/psv052}, 67, 64

\bibitem[\protect\citeauthoryear{Iwamoto, Amano, Hoshino  \& Matsumoto}{Iwamoto
  et~al.}{2017}]{Iwamoto2017}
Iwamoto M.,  Amano T.,  Hoshino M.,   Matsumoto Y.,  2017, \mn@doi [\apj]
  {10.3847/1538-4357/aa6d6f}, 840, 52

\bibitem[\protect\citeauthoryear{Iwamoto, Amano, Matsumoto, Matsukiyo  \&
  Hoshino}{Iwamoto et~al.}{2022}]{Iwamoto2022}
Iwamoto M.,  Amano T.,  Matsumoto Y.,  Matsukiyo S.,   Hoshino M.,  2022,
  \mn@doi [\apj] {10.3847/1538-4357/ac38aa}, 924, 108

\bibitem[\protect\citeauthoryear{Kaw \& Dawson}{Kaw \& Dawson}{1970}]{Kaw1970}
Kaw P.,  Dawson J.,  1970, \mn@doi [Phys. Fluids] {10.1063/1.1692942}, 13, 472

\bibitem[\protect\citeauthoryear{Kaw, Schmid  \& Wilcox}{Kaw
  et~al.}{1973}]{Kaw1973}
Kaw P.~K.,  Schmid G.,   Wilcox T.,  1973, \mn@doi [Phys. Fluids]
  {10.1063/1.1694552}, 16, 1522

\bibitem[\protect\citeauthoryear{Kruer}{Kruer}{1988}]{Kruer1988}
Kruer W.~L.,  1988, The Physics of Laser Plasma Interactions.
Addison-Wesley, Boston

\bibitem[\protect\citeauthoryear{Lorimer, Bailes, McLaughlin, Narkevic  \&
  Crawford}{Lorimer et~al.}{2007}]{Lorimer2007}
Lorimer D.~R.,  Bailes M.,  McLaughlin M.~A.,  Narkevic D.~J.,   Crawford F.,
  2007, \mn@doi [Science] {10.1126/science.1147532}, 318, 777

\bibitem[\protect\citeauthoryear{Luan \& Goldreich}{Luan \&
  Goldreich}{2014}]{Luan2014}
Luan J.,  Goldreich P.,  2014, \mn@doi [\apjl] {10.1088/2041-8205/785/2/L26},
  785, L26

\bibitem[\protect\citeauthoryear{Luo et~al.,}{Luo et~al.}{2020}]{Luo2020}
Luo R.,  et~al., 2020, \mn@doi [Nature] {10.1038/s41586-020-2827-2}, 586, 693

\bibitem[\protect\citeauthoryear{Lyubarsky}{Lyubarsky}{2014}]{Lyubarsky2014}
Lyubarsky Y.,  2014, \mn@doi [\mnras] {10.1093/mnrasl/slu046}, 442, L9

\bibitem[\protect\citeauthoryear{Lyubarsky}{Lyubarsky}{2021}]{Lyubarsky2021}
Lyubarsky Y.,  2021, \mn@doi [Universe] {10.3390/universe7030056}, 7, 56

\bibitem[\protect\citeauthoryear{Margalit, Metzger  \& Sironi}{Margalit
  et~al.}{2020a}]{Margalit2020a}
Margalit B.,  Metzger B.~D.,   Sironi L.,  2020a, \mn@doi [\mnras]
  {10.1093/mnras/staa1036}, 494, 4627

\bibitem[\protect\citeauthoryear{Margalit, Beniamini, Sridhar  \&
  Metzger}{Margalit et~al.}{2020b}]{Margalit2020b}
Margalit B.,  Beniamini P.,  Sridhar N.,   Metzger B.~D.,  2020b, \mn@doi
  [\apjl] {10.3847/2041-8213/abac57}, 899, L27

\bibitem[\protect\citeauthoryear{Matsukiyo \& Hada}{Matsukiyo \&
  Hada}{2003}]{Matsukiyo2003}
Matsukiyo S.,  Hada T.,  2003, \mn@doi [Phys. Rev. E]
  {10.1103/PhysRevE.67.046406}, 67, 046406

\bibitem[\protect\citeauthoryear{Matsumoto, Amano, Kato  \& Hoshino}{Matsumoto
  et~al.}{2015}]{Matsumoto2015}
Matsumoto Y.,  Amano T.,  Kato T.~N.,   Hoshino M.,  2015, \mn@doi [Sci]
  {10.1126/science.1260168}, 347, 974

\bibitem[\protect\citeauthoryear{Matsumoto, Amano, Kato  \& Hoshino}{Matsumoto
  et~al.}{2017}]{Matsumoto2017}
Matsumoto Y.,  Amano T.,  Kato T.~N.,   Hoshino M.,  2017, \mn@doi [\prl]
  {10.1103/PhysRevLett.119.105101}, 119, 105101

\bibitem[\protect\citeauthoryear{Max}{Max}{1973a}]{Max1973b}
Max C.~E.,  1973a, \mn@doi [Phys. Fluids] {10.1063/1.1694509}, 16, 1277

\bibitem[\protect\citeauthoryear{Max}{Max}{1973b}]{Max1973a}
Max C.~E.,  1973b, \mn@doi [Phys. Fluids] {10.1063/1.1694545}, 16, 1480

\bibitem[\protect\citeauthoryear{Max, Arons  \& Langdon}{Max
  et~al.}{1974}]{Max1974}
Max C.~E.,  Arons J.,   Langdon A.~B.,  1974, \mn@doi [\prl]
  {10.1103/PhysRevLett.33.209}, 33, 209

\bibitem[\protect\citeauthoryear{Metzger, Margalit  \& Sironi}{Metzger
  et~al.}{2019}]{Metzger2019}
Metzger B.~D.,  Margalit B.,   Sironi L.,  2019, \mn@doi [\mnras]
  {10.1093/mnras/stz700}, 485, 4091

\bibitem[\protect\citeauthoryear{Michilli et~al.,}{Michilli
  et~al.}{2018}]{Michilli2018}
Michilli D.,  et~al., 2018, \mn@doi [Nature] {10.1038/nature25149}, 553, 182

\bibitem[\protect\citeauthoryear{Mima \& Nishikawa}{Mima \&
  Nishikawa}{1975}]{Mima1975}
Mima K.,  Nishikawa K.,  1975, \mn@doi [J. Phys. Soc. Jpn.]
  {10.1143/JPSJ.38.1742}, 38, 1742

\bibitem[\protect\citeauthoryear{Mima \& Nishikawa}{Mima \&
  Nishikawa}{1984}]{Mima1984}
Mima K.,  Nishikawa K.,  1984, Basic Plasma Physics.
North-Holland Publishing Company, Amsterdam

\bibitem[\protect\citeauthoryear{Nimmo et~al.,}{Nimmo et~al.}{2021}]{Nimmo2021}
Nimmo K.,  et~al., 2021, \mn@doi [Nature Astronomy]
  {10.1038/s41550-021-01321-3}, 5, 594

\bibitem[\protect\citeauthoryear{Plotnikov \& Sironi}{Plotnikov \&
  Sironi}{2019}]{Plotnikov2019}
Plotnikov I.,  Sironi L.,  2019, \mn@doi [\mnras] {10.1093/mnras/stz640}, 485,
  3816

\bibitem[\protect\citeauthoryear{Plotnikov, Grassi  \& Grech}{Plotnikov
  et~al.}{2018}]{Plotnikov2018}
Plotnikov I.,  Grassi A.,   Grech M.,  2018, \mn@doi [\mnras]
  {10.1093/mnras/sty979}, 477, 5238

\bibitem[\protect\citeauthoryear{Schluck, Lehmann  \& Spatschek}{Schluck
  et~al.}{2017}]{Schluck2017}
Schluck F.,  Lehmann G.,   Spatschek K.~H.,  2017, \mn@doi [Phys. Rev. E]
  {10.1103/PhysRevE.96.053204}, 96, 053204

\bibitem[\protect\citeauthoryear{Sironi, Plotnikov, N{\"{a}}ttil{\"{a}}  \&
  Beloborodov}{Sironi et~al.}{2021}]{Sironi2021}
Sironi L.,  Plotnikov I.,  N{\"{a}}ttil{\"{a}} J.,   Beloborodov A.~M.,  2021,
  \mn@doi [\prl] {10.1103/PhysRevLett.127.035101}, 127, 035101

\bibitem[\protect\citeauthoryear{Sluijter \& Montgomery}{Sluijter \&
  Montgomery}{1965}]{Sluijter1965}
Sluijter F.~W.,  Montgomery D.,  1965, \mn@doi [Phys. Fluids]
  {10.1063/1.1761263}, 8, 551

\bibitem[\protect\citeauthoryear{{Sobacchi}, {Lyubarsky}, {Beloborodov}  \&
  {Sironi}}{{Sobacchi} et~al.}{2021}]{Sobacchi2020}
{Sobacchi} E.,  {Lyubarsky} Y.,  {Beloborodov} A.~M.,   {Sironi} L.,  2021,
  \mn@doi [\mnras] {10.1093/mnras/staa3248}, 500, 272

\bibitem[\protect\citeauthoryear{Sobacchi, Lyubarsky, Beloborodov  \&
  Sironi}{Sobacchi et~al.}{2022}]{Sobacchi2022a}
Sobacchi E.,  Lyubarsky Y.,  Beloborodov A.~M.,   Sironi L.,  2022, \mn@doi
  [\mnras] {10.1093/mnras/stac251}, 511, 4766

\bibitem[\protect\citeauthoryear{{Sobacchi}, {Lyubarsky}, {Beloborodov},
  {Sironi}  \& {Iwamoto}}{{Sobacchi} et~al.}{2023}]{Sobacchi2022b}
{Sobacchi} E.,  {Lyubarsky} Y.,  {Beloborodov} A.~M.,  {Sironi} L.,   {Iwamoto}
  M.,  2023, \mn@doi [\apjl] {10.3847/2041-8213/acb260}, 943, L21

\bibitem[\protect\citeauthoryear{Tabak et~al.,}{Tabak et~al.}{1994}]{Tabak1994}
Tabak M.,  et~al., 1994, \mn@doi [Phys. Plasmas] {10.1063/1.870664}, 1, 1626

\bibitem[\protect\citeauthoryear{Tajima \& Dawson}{Tajima \&
  Dawson}{1979}]{Tajima1979}
Tajima T.,  Dawson J.~M.,  1979, \mn@doi [\prl] {10.1103/PhysRevLett.43.267},
  43, 267

\makeatother
\end{thebibliography}

\appendix

\section{Stimulated Brillouin Scattering}\label{app:sbs}

\begin{figure*}
  \begin{center}
    \includegraphics[width=8cm]{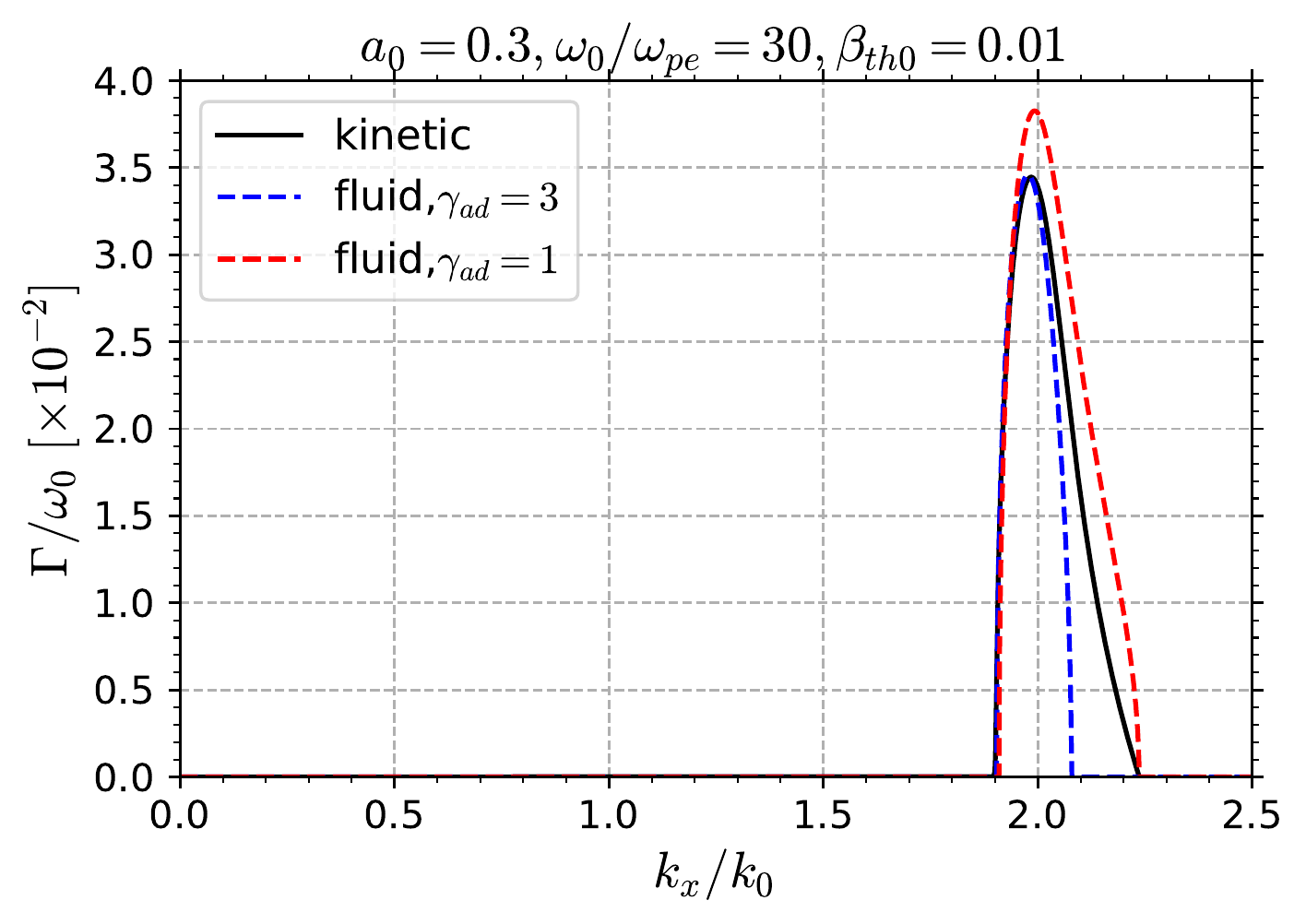}
    \includegraphics[width=8cm]{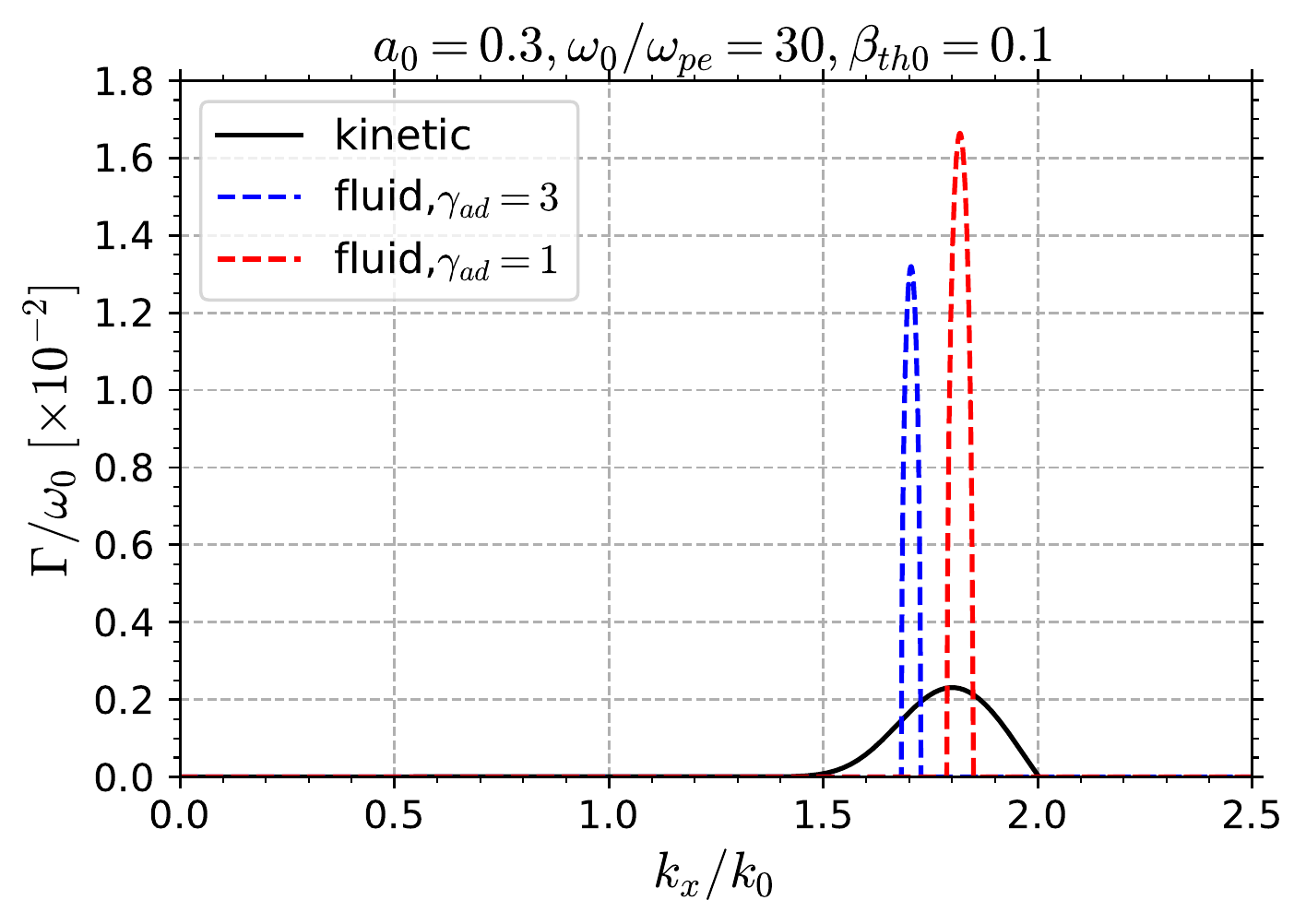}
  \end{center}
\caption{Linear growth rate of the SBS for strong
(left) and weak (right) coupling cases.}
\label{fig:sbs-lin}
\end{figure*}

The maximum growth rate of the SBS can be derived from the
same dispersion relation as the FI.
$\cos\theta_\pm=-1$ is satisfied for the backward
scattering and $D_+$ is non-resonant for the SBS.
The dispersion relation \ref{eq:gamf} reduces to
\begin{equation}
  (\omega^2-c_s^2k^2)\left[\omega+\frac{c^2}{2\omega_0}(k-2k_0)k\right]
  =-\frac{a_0^2\omega_{pe}^2c^2k^2}{4\omega_0}
\end{equation}
Substituting $\omega = c_sk+i\Gamma$,
where $\Gamma \ll c_sk$ for the weak coupling,
we obtain
\begin{equation}
  \Gamma^2-\frac{a_0^2\omega_{pe}^2c^2k}{8c_s\omega_0}-\frac{i\Gamma c^2}{2\omega_0}
  \left[k-2\left(k_0-\frac{c_s\omega_0}{c^2}\right)\right]k = 0.
\end{equation}
The maximum growth rate and corresponding wavevector are written as
\begin{gather}
  \Gamma_{max}^{SBS} = \frac{a_0\omega_{pe}}{2\sqrt{\beta_s}}, \\
  k_x^{SBS} = 2(1-\beta_s)k_0.
\end{gather}
The validity condition $\Gamma \ll c_sk$
now becomes
\begin{equation}
  \label{eq:sbsw}
  \beta_s \gg \left( a_0\frac{\omega_{pe}}{\omega_0}\right)^{\frac{2}{3}}.
\end{equation}
Here we have neglected factors of order of unity.
For the strong coupling $\Gamma \gg c_sk$, we obtain
\begin{equation}
  \omega^3+\frac{c^2}{2\omega_0}(k-2k_0)k\omega^2
  +\frac{a_0^2\omega_{pe}^2c^2k^2}{4\omega_0}=0
\end{equation}
The growth rate takes its maximum at around $k = 2k_0$ and we find
\begin{equation}
  \omega = (a_0^2\omega_{pe}^2\omega_0)^{\frac{1}{3}}e^{\frac{\pi i}{3}}.
\end{equation}
We finally obtain
\begin{gather}
  \Gamma_{max}^{SBS} =\frac{\sqrt{3}}{2}(a_0^2\omega_{pe}^2\omega_0)^{\frac{1}{3}},\\
  k_x^{SBS} = 2k_0,
\end{gather}
and the validity condition is
\begin{equation}
  \label{eq:sbss}
  \beta_s \ll \left( a_0\frac{\omega_{pe}}{\omega_0}\right)^{\frac{2}{3}}.
\end{equation}
Our parameters $(a_0,\omega_0/\omega_{pe},\beta_{th0})=(0.3,30,0.1)$
and $(0.3,30,0.01)$ satisfy the weak (Equation \ref{eq:sbsw}) and 
strong (Equation \ref{eq:sbss}) coupling conditions for the SBS, respectively.
We numerically derive the linear growth rate of the SBS and show it in Figure
\ref{fig:sbs-lin} for the strong (left) and weak (right) coupling cases.
The SBS grows faster than the FI for both cases.
For the weak coupling case, the kinetic growth rate is much smaller
than the fluid one because the fluid growth rate is always overestimated due
to the absence of  Landau damping.
The unstable wavevector is smaller than $2k_0$, indicating that
the backscattered waves propagating the $-x$ direction
are not resolved in our simulation box
$L_x \times L_y = \lambda_0 \times 120 \lambda_0$,
which helps to suppress the SBS as already discussed by \cite{Ghosh2022}. In fact, 
for hot plasmas with $\beta_{th0}=0.1$, we find that the amplitude of the SBS-generated density
fluctuations is vary small ($\delta n_e/n_0 \sim 10^{-2}$), i.e., the SBS is suppressed.

\begin{figure}
  \begin{center}
  \includegraphics[width=8.5cm]{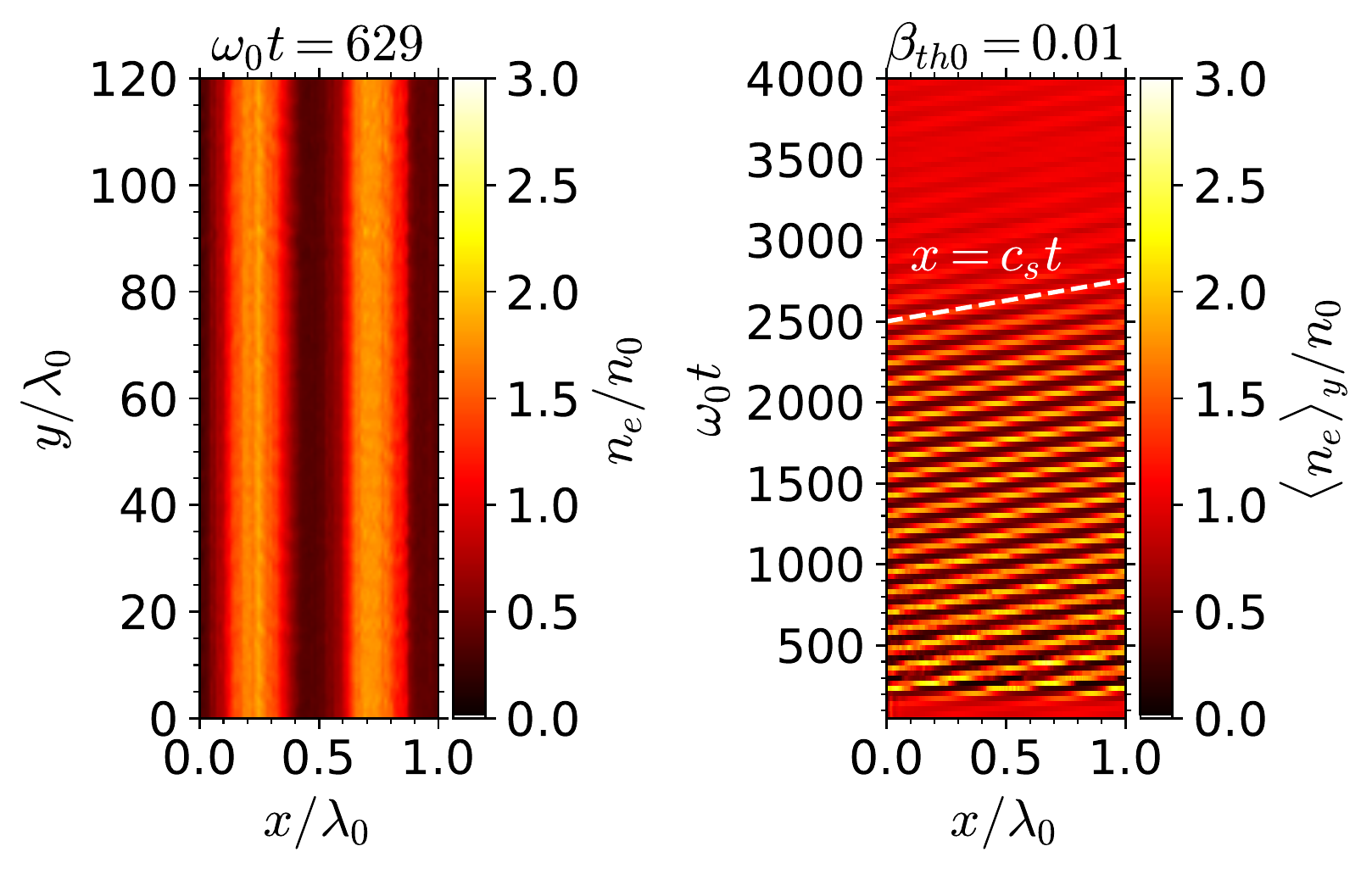}
  \end{center}
\caption{Left: snapshot of the electron density at $\omega_0t=629$.
Right: time evolution of the $y$-averaged electron density.
The white dashed line represents fluctuations propagating with
the sound speed $c_s$.}
\label{fig:nsbs}
\end{figure}

In contrast, in the strong coupling case the unstable wavevector can exceed $2k_0$ 
and thus SBS operates even for our simulation setting.
Figure \ref{fig:nsbs} shows the snapshot of the electron density at
$\omega_0t=629$ for $\beta_{th0} = 0.01$ (left).
The density fluctuation at the wavenumber $k_x \sim 2k_0$ is clearly seen, 
indicating that the SBS indeed operates for the strong coupling regime.
The time evolution of the $y$-averaged electron density
is shown in the right panel of Figure \ref{fig:nsbs}.
The white dashed line represents fluctuations propagating with
the sound speed $c_s$, where the adiabatic index $\gamma_{ad}=3$ is
assumed, showing that the the density fluctuation at $k_x \sim 2k_0$ is
propagating in the $+\bvec{x}$ direction with the sound speed.
Since the SBS is expected to generate the forward-propagating sound-like
waves, this provides a clear proof of the SBS.

\begin{figure}
  \begin{center}
  \includegraphics[width=8cm]{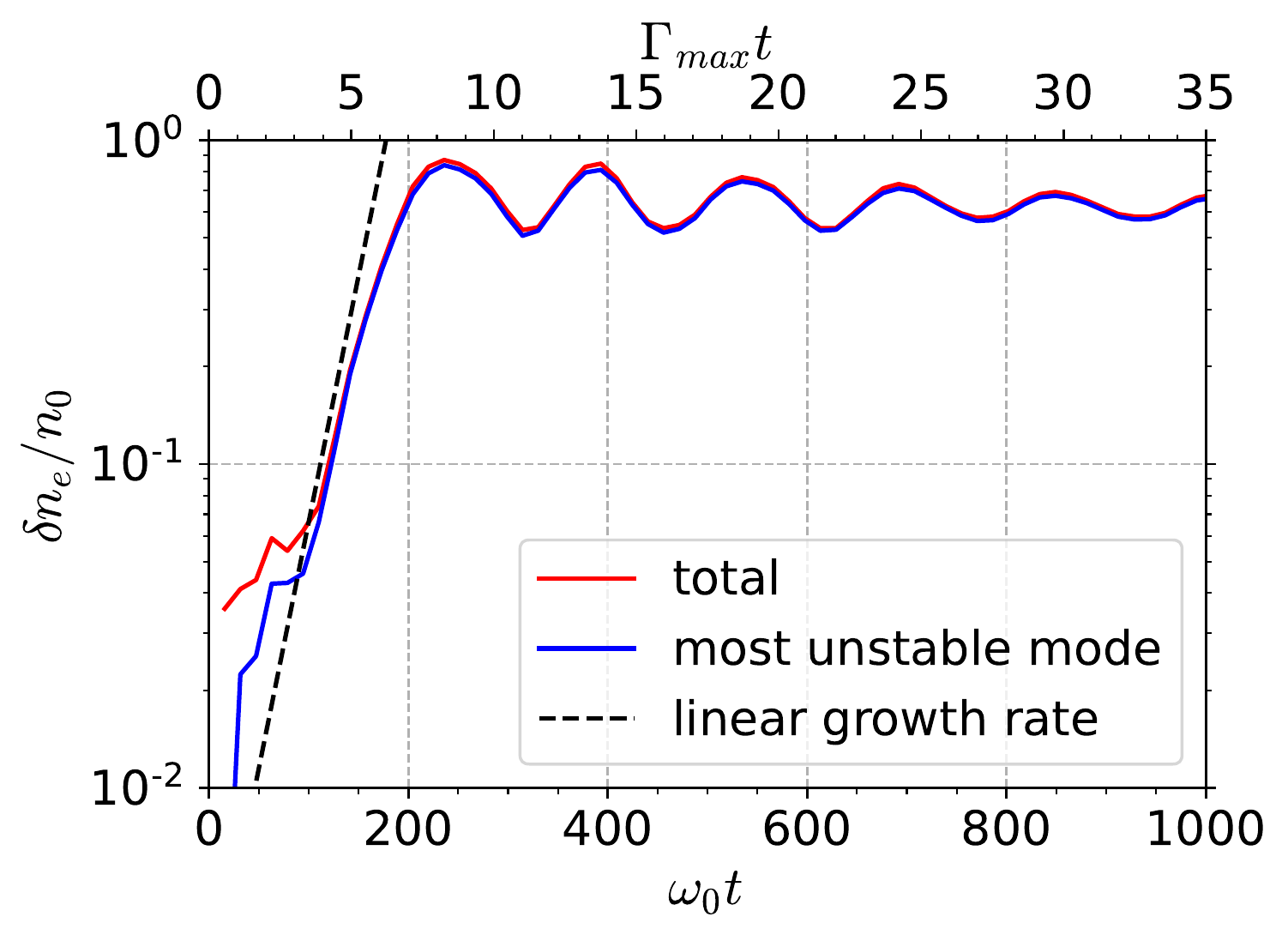}
  \end{center}
\caption{Time evolution of the SBS. The amplitude is calculated from
the Fourier components of the $y$-averaged density fluctuations for
$\beta_{th0}=0.01$.
The most unstable modes (blue) and integral of all modes (red) are
shown. The black dashed lines correspond to $\propto e^{\Gamma_{max}t}$,
where $\Gamma_{max}$ is determined from the linear theory 
(Equation \ref{eq:gamk} for $k_y=0$).}
\label{fig:sbs-growth}
\end{figure}

Figure \ref{fig:sbs-growth} shows the
time evolution of Fourier components of the $y$-averaged density fluctuations
for the strong coupling case $\beta_{th0}=0.01$.
The black dashed lines are maximum growth rates $\Gamma_{max}$ of the SBS
determined from the linear theory (Equation \ref{eq:gamk} for $k_y=0$), showing a
good agreement with our simulation result.
Based on the above analysis,
we conclude that the longitudinal density fluctuation
with $k_x \sim 2k_0$ originates from the SBS
and that the SBS is not fully suppressed by our numerical setting
for the strong coupling case.

\section{Out-of-plane Vector Potential}\label{app:out}

In the main text, we focus on the pump wave vector potential lying in
the $y$ direction. One can choose the out-of-plane vector potential
($z$ direction in our coordinates) and the corresponding wave fields are
\begin{align}
  \bvec{E_0} &= (0,0,E_0\cos{k_0x}), \\
  \bvec{B_0} &= \left(0,\frac{ck_0}{\omega_0}E_0\cos{k_0x}, 0 \right).
\end{align}
In this case, $\bvec{A_0} \parallel \bvec{\delta A_{\pm}}$ (i.e, $\cos\theta_{\pm}=1$)
is always satisfied regardless of the scattering direction, and
thus side scattering ($\bvec{k_0} \perp \bvec{k_{\pm}}$)
survives unlike in the in-plane configuration.

\begin{figure}
  \begin{center}
    \includegraphics[width=8cm]{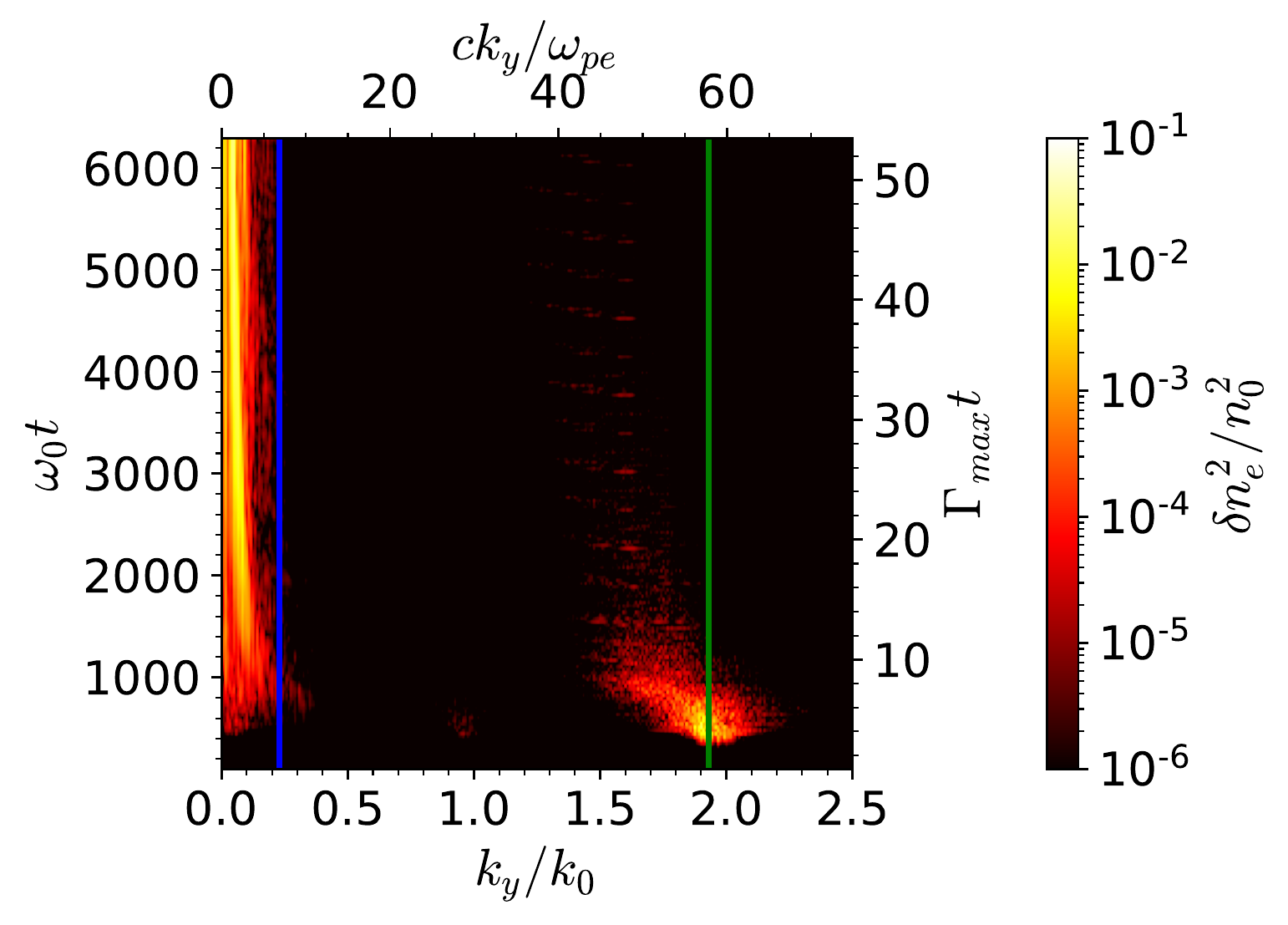}
  \end{center}
\caption{Time evolution of the power spectrum of the $x$-averaged
electron density fluctuations for $\beta_{th0}=0.01$ with the out-of-plane
vector potential. The blue and green lines correspond to
the most unstable mode of the FI and SBS, respectively.}
\label{fig:fourier_out}
\end{figure}

Figure \ref{fig:fourier_out} shows the time evolution of the $x$-averaged power spectrum
for $\beta_{th0}=0.01$ in the out-of-plane configuration.
The numerical parameters are identical to the in-plane configuration in the
main text and only the direction of the initial vector potential changes.
The clear peak can be no longer seen near the theoretical most unstable mode
of the FI (the blue line in Figure \ref{fig:fourier_out}).
The filaments merge much earlier than the in-plane configuration.
Furthermore, the mode with $k_y \sim 2k_0$ apparently grows faster than
others, which is not observed the in-plane configuration.
We think that the side-scattered wave plays the role of a pump wave 
and the peak at $k_y \sim 2k_0$ can be attributed to a secondary SBS of the side-scattered 
wave. In fact, the green line indicates the most unstable mode of the secondary SBS,
showing a good agreement with the observed peak.
Although the secondary SBS may induce the side-scattered wave again,
the wavevector is almost identical to the pump wave and these waves
cannot be distinguished.
We here assumed that the wavenumber of the side-scattered wave,
which plays a role for the pump wave of the secondary SBS, satisfies
$\bvec{k_s} \simeq \pm k_0 \bvec{\hat{y}}$.
We confirmed this for $\beta_{th0}=0.1$ and the secondary SBS works
for both strong and weak coupling cases.
Note that our simulation setting can numerically suppress only the
back-scattering which is the dominant mode of the SBS
\citep{Ghosh2022}. The side-scattering survives even for the weak coupling 
regime in which the backward SBS is well-suppressed.

\begin{figure}
  \begin{center}
    \includegraphics[width=8cm]{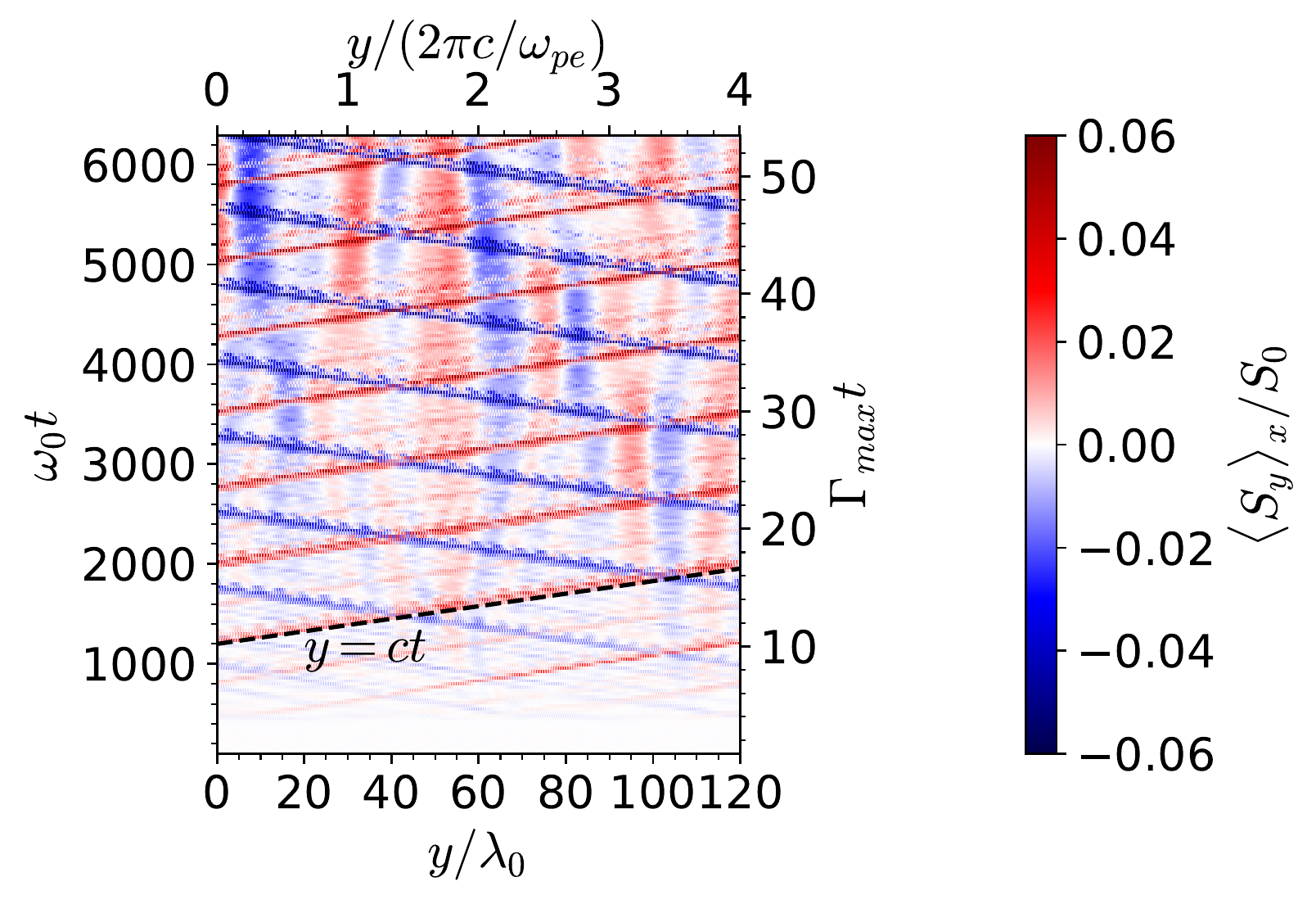}
  \end{center}
\caption{Time evolution of the $y$ component of the $x$-averaged Poynting flux
for $\beta_{th0}=0.01$ with the out-of-plane vector potential.
The black dashed line indicates
the electromagnetic waves propagating in the $y$ direction.}
\label{fig:sy}
\end{figure}

Figure \ref{fig:sy} shows the time evolution of the $y$ component of the
$x$-averaged Poynting flux $\langle S_y \rangle_x$ for $\beta_{th0}=0.01$,
where $\langle S_y \rangle_x$ is normalized by the initial mean flux
$S_0=E_0^2/8\pi$. The grid-like structures are clearly seen in addition to
the transverse filamentary structures from the FI. The black dashed line
represents the electromagnetic waves propagating in the $y$ direction,
indicating that the grid-like structures originate from side-scattered
waves traveling toward the $\pm y$ direction. It has been argued that
side scattering for the out-of-plane vector potentials is numerically enhanced 
due to the periodic boundary condition in the $y$ direction 
\cite[e.g.,][]{Cohen2005}. We find that the side scattering preferentially works 
and dominates over the FI for the out-of-plane vector potentials.

\section{Numerical Convergence}\label{app:conv}

Here, we demonstrate the convergence of the growth rate and saturation level
with respect to the number of particles per cell per species $n_0 \Delta x^2$.

\begin{figure}
  \begin{center}
    \includegraphics[width=8cm]{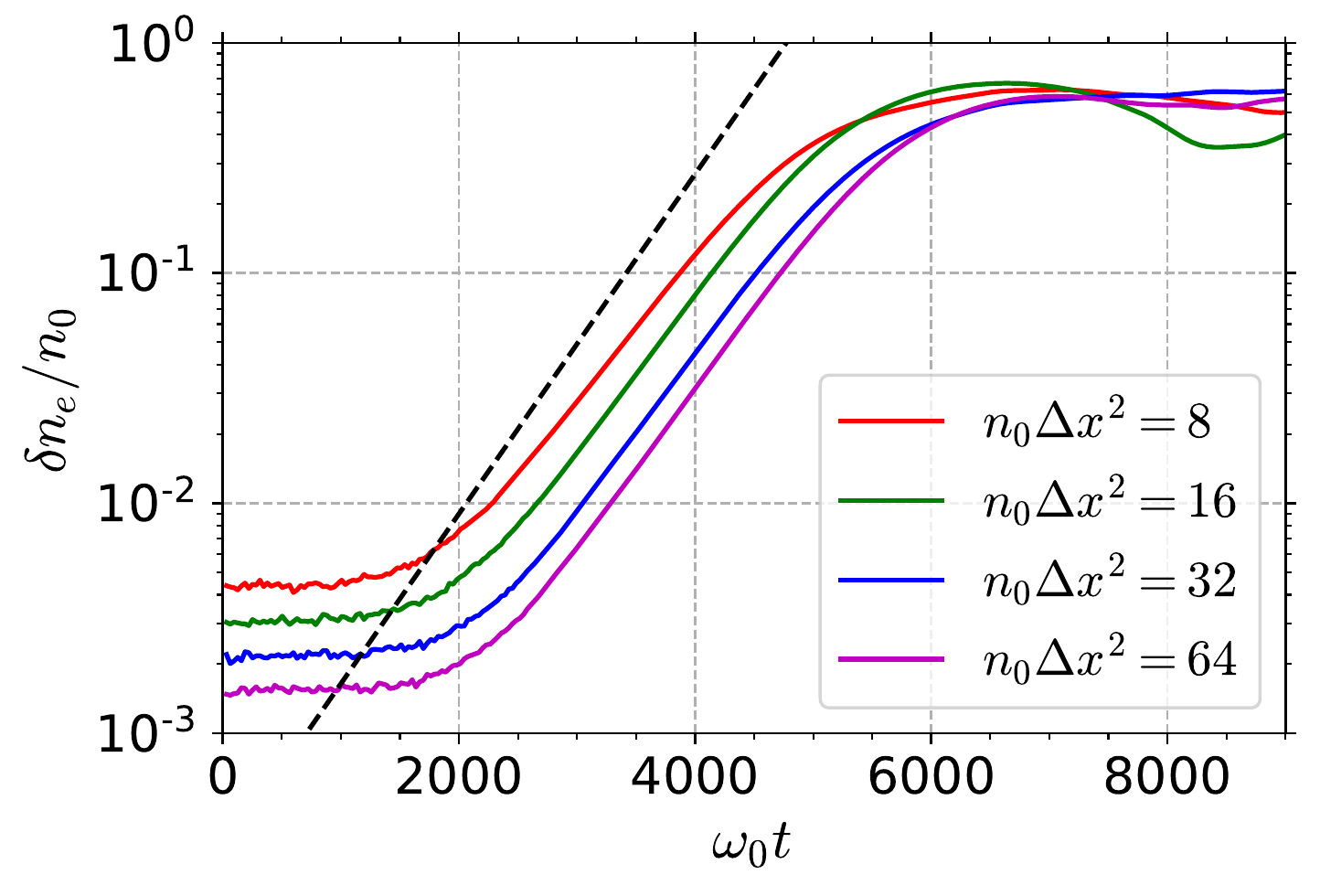}
  \end{center}
\caption{Numerical convergence with respect to the number of particles per cell 
per species for $\beta_{th0}=0.1$. The total of all Fourier modes of the 
transverse electron density fluctuations $\delta n_e$ is shown for 
$n_0\Delta x^2=8$ (red), $16$ (green), $32$ (blue), and $64$ (purple). 
The black dashed lines represent $\propto e^{\Gamma_{max}t}$.}
\label{fig:conv}
\end{figure}

Figure \ref{fig:conv} shows the time evolution of the spectrum-integrated signal
of $\delta n_e (y)$ for $\beta_{th0}=0.1$ for
$n_0\Delta x^2=8$ (red), $16$ (green), $32$ (blue), and $64$ (purple).
The black dashed line represents the fastest-growing mode from the linear theory.
It is natural that the initial noise level should decrease as $n_0\Delta x^2$ increases.
Both growth rate and saturation level converge for $n_0\Delta x^2 \geq 32$.
Based on this result, we choose $n_0\Delta x^2 = 32$ in the main text.
In fact, the blue line shown in Figure \ref{fig:conv} is the same as the red line in 
the right panel of Figure \ref{fig:growth}.

\bsp
\label{lastpage}
\end{document}